\documentclass[journal]{IEEEtran} 

\usepackage{amsmath, amsfonts, amssymb, mathtools}
\usepackage{graphicx} 
\usepackage{array}
\usepackage{textcomp}
\usepackage{stfloats}
\usepackage{url}
\usepackage{booktabs}
\usepackage{verbatim}
\usepackage{float} 
\usepackage{color}
\usepackage{cite}
\usepackage{bbm}
\usepackage{verbatim} 
\usepackage{bm}
\usepackage{algorithm}
\usepackage{algpseudocode}
\usepackage{booktabs}
\usepackage[font=footnotesize]{caption}

\usepackage{xcolor}
\definecolor{myblue}{RGB}{0,90,180}
\definecolor{mygreen}{RGB}{0,130,0}

\usepackage{graphicx}    
\usepackage{caption}     
\usepackage{subcaption}  

\usepackage{hyperref} 
\hypersetup{
	colorlinks=false,        
	pdfborder={0.5 0.5 0.5}  
}
\begin{document}

	\bstctlcite{IEEEexample:BSTcontrol}

	\title{QoS-Aware Hierarchical Reinforcement Learning for 
    Joint Link Selection and Trajectory Optimization in SAGIN-Supported UAV Mobility Management}

	\author{ 
		Jiayang Wan, \IEEEmembership{Graduate Student Member, IEEE}, 
		Ke He, \IEEEmembership{Member, IEEE}, 
		\\
Yafei Wang, \IEEEmembership{Graduate Student Member, IEEE},
		Fan Liu, \IEEEmembership{Senior Member, IEEE},
				\\
        Wenjin Wang, \IEEEmembership{Member, IEEE},
		Shi Jin, \IEEEmembership{Fellow, IEEE}

        \thanks{
		 Jiayang Wan, Yafei Wang, and Wenjin Wang are with the National Mobile Communications Research
		Laboratory, Southeast University, Nanjing 210096, China, and also
		with Purple Mountain Laboratories, Nanjing 211100, China (e-mail:
        \{jywan, wangyf, wangwj\}@seu.edu.cn).} 
			\thanks{
			K. He is with the Interdisciplinary Centre for Security, Reliability and Trust (SnT), University
of Luxembourg, L-1855 Luxembourg, Luxembourg (e-mail: ke.he@ieee.org).} 
\thanks{
F. Liu and S. Jin are with the National Mobile Communications Research Laboratory, Southeast University, Nanjing 210096, China (e-mail:
 \{fan.liu, jinshi\}@seu.edu.cn).}
	}

	\maketitle

	\begin{abstract}
Due to the significant variations in unmanned aerial vehicle (UAV) altitude and horizontal mobility,  it becomes difficult for any single network to ensure continuous and reliable three-dimensional coverage.  
Towards that end, the space-air-ground integrated network (SAGIN) has emerged as an essential architecture for enabling ubiquitous UAV connectivity.
        To address the pronounced disparities in coverage and signal characteristics across heterogeneous networks, 
        this paper formulates UAV mobility management in SAGIN as a constrained multi-objective joint optimization problem. The formulation couples discrete link selection with continuous trajectory optimization.
	Building on this, we propose a two-level multi-agent hierarchical deep reinforcement learning (HDRL) framework that decomposes the problem into two alternately solvable subproblems.
		To map complex link selection decisions into a compact discrete action space, we conceive a double deep Q-network (DDQN) algorithm in the top-level, which achieves stable and high-quality policy learning through double Q-value estimation.
		To handle the continuous trajectory action space while satisfying  quality of service (QoS) constraints,  we integrate the maximum-entropy mechanism of the soft actor-critic (SAC) and employ a Lagrangian-based constrained SAC (CSAC) algorithm in the lower-level  that dynamically adjusts the Lagrange multipliers to balance constraint satisfaction and policy optimization.
		Moreover, the proposed algorithm can be extended to multi-UAV scenarios under the centralized training and decentralized execution (CTDE) paradigm, which enables more generalizable policies.
	Simulation results demonstrate that the proposed scheme substantially outperforms existing benchmarks in throughput, link switching frequency and QoS satisfaction.
	\end{abstract}
	
	
	\vspace{-4mm}
	\section{Introduction}\label{Introduction}
	\vspace{-1mm}

	\IEEEPARstart{W}{i}th
	the rapid expansion of the low-altitude economy, existing terrestrial networks (TNs) \cite{saad2019vision} cannot ensure continuous and reliable coverage in low-altitude airspace characterized by significant altitude variations and high horizontal mobility, resulting in pronounced coverage gaps.
	To overcome this limitation, emerging non-terrestrial networks (NTNs), such as low-altitude wireless networks (LAWNs) and satellite communication systems \cite{wu2025distributed, wang2025statistical}, serve as effective complements by providing wide-area connectivity in regions beyond the coverage of TNs.
	Consequently, the space-air-ground integrated network (SAGIN) \cite{yang20196g,dang2020should}, which integrates TNs with LAWNs and satellite communication networks, has emerged as a key architecture for enabling wide-area and seamless connectivity in low-altitude airspace \cite{xu2023space,hou2024beam}.
	
	In SAGIN, unmanned aerial vehicles (UAVs), as a new class of aerial users, are increasingly deployed in mission-critical applications such as environmental monitoring, aerial imaging, air transportation, and disaster relief \cite{liufan}.
    During operation, UAVs must simultaneously perform link selection and trajectory optimization.
However, due to the coexistence of heterogeneous networks with diverse coverage, frequent link switching is triggered, resulting in excessive overhead and unstable connectivity \cite{Khosravi2021}.
	Moreover, since the distinct link dynamics between TNs and NTNs, where TN links vary rapidly with propagation distance while NTN links remain relatively stable, UAV trajectory optimization becomes particularly challenging \cite{Haghrah2023}.

	\vspace{-5pt}
	
		\vspace{-2mm}
	
	\subsection{Prior Work}
Owing to significant variations in UAV altitude and high horizontal mobility, UAVs frequently traverse ground, aerial, and satellite coverage regions, making link selection a key research problem in UAV communications.
Depending on whether the decision is made within a single network or across multiple networks, link selection can be classified into two categories: intra-network link selection \cite{su2019novel,zhang2024toward,jaffry2020comprehensive,ho2020joint,karmakar2022mobility,yang2023dqn,kang2023joint,wang2022seamless} and inter-network link selection \cite{xu2017modeling,liu2023user,wang2024sustainable}.
Intra-network link selection primarily concerns UAV associations among nodes within the same network type.
In TNs, a UAV switches its link between ground base stations (GBSs) when a target GBS provides better channel quality than the serving GBS.
To reduce link switching frequency, link selection triggering is regulated via control parameters such as hysteresis, time-to-trigger (TTT), and cell individual offset (CIO) \cite{su2019novel,zhang2024toward}.
As emphasized in \cite{jaffry2020comprehensive}, efficient connectivity management is crucial for mobile nodes within the network, such as UAVs and trains.
The authors in \cite{ho2020joint} investigate a deep reinforcement learning (DRL)-based connectivity management scheme to dynamically adjust link selection decisions.
In \cite{karmakar2022mobility}, the authors study machine learning-based link selection prediction and demonstrate significant improvements in communication efficiency and reliability.
In contrast, connectivity management in NTNs exhibits relatively smooth link dynamics, where conventional TN-based criteria may trigger unnecessary link switching and excessive overhead \cite{yang2023dqn,kang2023joint,wang2022seamless}.
To address this issue, NTN-specific strategies have been proposed that incorporate satellite positions and predictable orbital information into link selection conditions, thereby reducing link switching frequency \cite{yang2023dqn}.
In addition, several studies have proposed multi-metric optimization approaches that jointly optimize resource allocation and traffic offloading to enhance NTN service capability \cite{kang2023joint}.
Moreover, a conditional handover (CHO)-based optimization strategy is proposed in \cite{wang2022seamless} for low Earth orbit (LEO) satellite networks, where a service continuity model is leveraged to effectively reduce link switching frequency.

When it comes to inter-network link selection, the substantial differences in signal characteristics and coverage footprints between TNs and NTNs render link selection more complex than its intra-network counterpart \cite{xu2017modeling,liu2023user,wang2024sustainable}.
A cross-level link selection model based on stochastic geometry is developed in \cite{xu2017modeling}, and closed-form expressions for key performance metrics, such as the link switching rate, failure rate, and ping-pong rate, are derived.
In \cite{liu2023user}, TN-NTN cross-domain link selection mechanisms are examined, and it is demonstrated that effective inter-network coordination can enhance resource utilization for remote users.
In addition, a graph-based UAV mobility management strategy is proposed in \cite{wang2024sustainable}, where TNs and NTNs are treated as both competitive and cooperative network entities, and the inter-network link selection process is dynamically optimized to improve communication continuity while reducing link switching frequency.
However, these studies, while focusing on link selection optimization in heterogeneous networks, do not adequately account for UAV trajectory optimization, which significantly affects both link selection frequency and stability.

As a key enabling technique for UAV communication systems, trajectory optimization plays a critical role in enhancing communication performance and mission efficiency through intelligent flight path planning \cite{zhang2018cellular,bulut2018trajectory,zhang2019trajectory,zeng2021simultaneous,zhan2022energy,madelkhanova2022optimization,narmeen2025coordinated,du2025handover}.
Existing optimization- and heuristic-based approaches \cite{zhang2018cellular,bulut2018trajectory,zhang2019trajectory} mainly include graph-theoretic methods \cite{zhang2018cellular}, convex optimization \cite{bulut2018trajectory}, and dynamic programming (DP) \cite{zhang2019trajectory}.
Specifically, \cite{zhang2018cellular} develops a trajectory design framework that integrates graph theory with convex optimization, with the objective of minimizing the UAV’s flight time while maintaining continuous connectivity with at least one GBS.
Building upon this line of research, \cite{zhang2019trajectory} adopts a dynamic programming (DP) approach to generate UAV trajectories that effectively constrain the duration of disconnection from GBSs during task execution.
In addition, \cite{bulut2018trajectory} incorporates communication outage constraints into a graph-theoretic optimization framework to jointly optimize UAV trajectory and link reliability.
However, these methods are inherently designed for single-stage or short-horizon objectives and fail to capture long-term cumulative performance, rendering them less effective in complex and dynamic communication scenarios.

To address the limitations of optimization methods in dynamic environments, RL-based methods, especially DRL
\cite{zeng2021simultaneous,zhan2022energy,madelkhanova2022optimization,narmeen2025coordinated,du2025handover}, have received increasing attention.
In particular, \cite{zeng2021simultaneous} proposed a DRL-based coverage-aware navigation framework for cellular-connected UAVs to minimize mission time and communication outage, while \cite{zhan2022energy} employed a deep Q-network (DQN)-based scheme to jointly optimize mission duration, trajectory, and BS association for energy-efficient and reliable connectivity.
An RL-based framework is developed in \cite{madelkhanova2022optimization,narmeen2025coordinated} to dynamically adjust the CIO of aerial and ground BSs, 
thereby maximizing user capacity while reducing link selection failures and ping-pong effects.
To minimize the frequency of link switching, \cite{du2025handover} optimizes UAV trajectories with a graph theory-based approach under communication and mission constraints.
The aforementioned methods discretize the continuous trajectory action space of UAVs to simplify learning and decision-making, which limits fine-grained modeling of continuous motion and hinders the full potential of trajectory optimization. 
Moreover, these studies typically select the link with the maximum reference signal received power (RSRP) or signal-to-interference-plus-noise ratio (SINR), which can lead to high link switching frequency in heterogeneous network.

	\vspace{-3mm}
	\subsection{Motivation and Main Contributions}
	\vspace{-0.5mm}

Although significant progress has been made in UAV trajectory optimization and link selection, the joint optimization of these two aspects for UAVs in SAGIN still faces two major challenges.
First, the distinct signal characteristics across heterogeneous networks render conventional optimization methods less effective, necessitating novel algorithmic designs.
Second, the strong coupling between discrete link selection and continuous trajectory optimization significantly increases the complexity of the joint optimization problem.
Moreover, most existing works discretize the continuous trajectory action space, which limits fine-grained modeling of UAV motion, and employ greedy RSRP/SINR-based link selection strategies, leading to high link switching frequency.

Motivated by the above challenges, this paper formulates the joint link selection and trajectory optimization problem for UAVs in SAGIN as a constrained multi-objective optimization problem. 
The objective is to maximize the overall system throughput while minimizing the link switching frequency and UAV flight time, subject to communication quality-of-service (QoS) constraints.
The main contributions of this work are summarized as follows:
	\begin{itemize}
		\item  
To efficiently address the strong coupling between discrete link selection variables and continuous trajectory optimization variables, we propose a two-level multi-agent hierarchical deep reinforcement learning (HDRL) framework \cite{kulkarni2016hierarchical} for joint link selection and trajectory optimization. The proposed framework decomposes the original problem into two subproblems that are solved in an alternating manner.
In the top-level, the UAV-BS association indicator is updated to determine the serving BS, based on which the lower level further optimizes UAV trajectory control.
Unlike conventional offline optimization methods, the proposed HDRL framework can be extended to multi-UAV scenarios under the centralized training and decentralized execution (CTDE) paradigm, facilitating stable and generalizable policy learning.
		\item
In the top-level, to compactly represent complex UAV-BS association decisions, we employ a double deep Q-network (DDQN) that maps the hybrid association space to two semantic actions: \textit{Remain} and \textit{Switch}.
By leveraging double Q-value estimation to decouple action selection from action evaluation, the top-level agent effectively suppresses overestimation bias and achieves stable and fast-converging UAV-BS association policies.
		\item
In the lower-level, to avoid performance degradation caused by discretizing the UAV’s continuous trajectory action space, we model trajectory control in a continuous action space.
However, under this formulation, UAVs must simultaneously satisfy QoS constraints, which leads to complex nonlinear coupling between the state and decision variables.
To tackle this challenge, we introduce a Lagrangian-based constrained soft actor-critic (CSAC) framework based on the principles of constrained reinforcement learning (CRL).
This framework not only incorporates the maximum-entropy policy of SAC to enhance exploration and training stability, but also explicitly embeds constraint terms into the standard RL objective.
As a result, the agent can maximize long-term cumulative rewards while dynamically adjusting the Lagrange multipliers to satisfy constraints, thereby avoiding tedious reward shaping and hyperparameter tuning.
	\end{itemize}

\textit{Organization:} The remainder of this paper is organized as follows.
Section~\ref{System Model} presents the system model for UAVs in SAGIN.
Section~\ref{Uav Handover Optimization} formulates the joint UAV link selection and trajectory optimization problem.
Section~\ref{Proposed HDRL Algorithm} presents the proposed HDRL-based algorithm.
Section~\ref{Simulation Results} presents the simulation results, and Section~\ref{Conclusion} concludes the paper.

\textit{Notation:} Throughout this paper, bold uppercase letters denote matrices, and bold lowercase letters denote vectors. The operators $(\cdot)^T$, $(\cdot)^*$, and $(\cdot)^H$ represent the transpose, complex conjugate, and conjugate transpose, respectively. The function $\delta(\cdot)$ denotes the Dirac delta, which is widely used in signal processing, while $\|\cdot\|_F$ and $\|\cdot\|_2$ denote the Frobenius norm and the $\ell_2$ norm, respectively.

	\vspace{-4mm}
	\section{System Model}\label{System Model}
	\vspace{-1mm}

In this section, we first present the overall SAGIN architecture in Section~\ref{Network Model}.
Then, the signal model is described in detail in Section~\ref{Signal Model}.
Subsequently, the UAV trajectory model and the link selection model are introduced in Sections~\ref{UAV Trajectory Model} and~\ref{UAV Handover Model}, which lay the foundation for   problem formulation.

	\vspace{-5pt}
	\vspace{-5pt}
	\vspace{-5pt}
	\subsection{SAGIN Architecture}\label{Network Model}

\begin{figure}[t]
		\centering		\includegraphics[width= 3.5in]{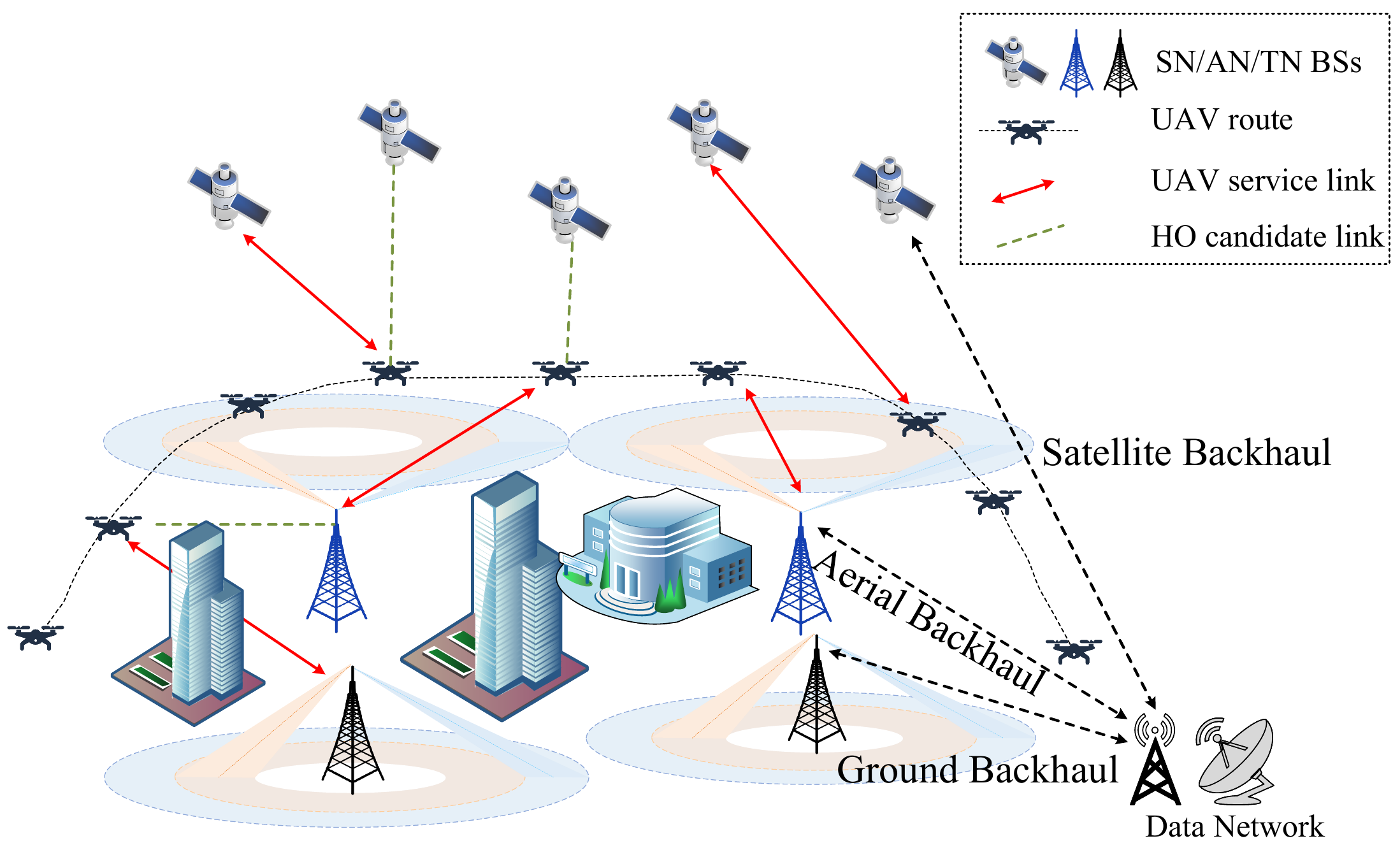}
		\caption{Illustration of UAV trajectory and link  selection in a SAGIN.}
		\label{fig1}
	\end{figure}

The considered SAGIN architecture is illustrated in Fig.~\ref{fig1} and consists of the following elements.

\subsubsection{Ground Network (GN)}
The ground segment comprises multiple communication systems, including but not limited to cellular networks and wireless local area networks (WLANs) \cite{conti2014mobile}.
In this paper, we consider a ground cellular network consisting of $G$ terrestrial cellular cells.
Each cell is served by a GBS site with three co-located BSs, denoted by $b^{\text{GN}} \in \mathcal{B}^{\text{GN}}$, each covering a $120^{\circ}$ sector \cite{3gpp36873}.

    \subsubsection{Air Network (AN)}
The aerial segment consists of UAVs, airships, balloon platforms, and exclusive-service aerial base stations (ABSs) for aerial users  \cite{kim2022non}, which collectively form LAWNs.
In this paper, we consider a low-altitude network consisting of $L$ aerial cells.
Each low-altitude cell is covered by one ABS site hosting three co-located BSs, denoted by $b^{\text{AN}} \in \mathcal{B}^{\text{AN}}$.
Each ABS employs an independently designed antenna tilt angle to ensure reliable and efficient communications in the low-altitude region \cite{kim2022non}.

\subsubsection{Space Network (SN)}
The space segment primarily comprises satellites operating at different altitudes and orbital configurations.
In this paper, we consider a space network consisting of $S$ low Earth orbit (LEO) satellites, collectively denoted by $\mathcal{B}^{\text{SN}}$.
Each satellite is equipped with multiple beams whose pointing directions remain fixed with respect to the satellite body.

\subsubsection{Gateway \& UAVs}
A gateway station integrates the service-area gateways of the ground, air, and space networks, establishing connections with all three network segments to facilitate content delivery from the Internet.
Meanwhile, a swarm of UAVs performs flight missions in the low-altitude region, flying at certain speeds and requiring continuous link selection and trajectory optimization to maintain seamless connectivity and enable data reception in SAGIN.
In what follows, we consider three types of links:
	\begin{itemize}
		\item \textit{GN-UAV links}, connecting a GBS $b^{\text{GN}}$ and a UAV $u$.
		\item \textit{SN-UAV links}, connecting a  satellite $b^{\text{SN}}$ and a UAV $u$.
		\item \textit{AN-UAV links}, connecting a ABS $b^{\text{AN}}$ and a UAV $u$.
	\end{itemize}
	For GN-UAV and AN-UAV links, all BSs are assumed to be deployed at the same altitude, 
	such that each BS provides coverage over a specific service area accessible to UAVs.
For SN-UAV links, owing to the rapid movement of LEO satellites, we consider $S$ LEO satellites to ensure continuous coverage along the UAV flight trajectory, thereby guaranteeing that at each time instant at least one satellite maintains connectivity with the UAV.
In this work, we assume that each UAV can be connected to only one BS at a time.

	\vspace{-5pt}
	\vspace{-5pt}
	
	\subsection{Signal Model}\label{Signal Model}

Let $h_{b,u}(t)$ denote the baseband-equivalent channel between the $b$-th BS and the $u$-th UAV at time~$t$, where $b \in \mathcal{B}$.
The BS set $\mathcal{B}$ is defined as $\mathcal{B} = \mathcal{B}^{\text{GN}} \cup \mathcal{B}^{\text{AN}} \cup \mathcal{B}^{\text{SN}}$, with $|\mathcal{B}| = B$.
Accordingly, the received signal power at time $t$ by the $u$-th UAV from the $b$-th BS is expressed as
	\vspace{-5pt}
	\begin{equation}
		\begin{aligned}
			p_{b,u}(t) 
			&= \bar{P}_{b} \big| h_{b,u}(t) \big|^2 \\
			&= \bar{P}_{b} \, \beta_{b,u}\!\big(\mathbf{q}_u(t)\big) 
			G_{b,u}\!\big(\mathbf{q}_u(t)\big) 
			\tilde{h}_{b,u}(t), 
			\quad b = 1, \cdots, B,
		\end{aligned}
		\label{eq:pb}
	\end{equation}
where $\bar{P}_b$ denotes the transmit power of the $b$-th BS, and $\beta_{b,u}(\cdot)$ and $G_{b,u}(\cdot)$ denote the large-scale channel gain and the BS antenna gain, respectively, both of which depend on the UAV location $\mathbf{q}_u(t)$. The term $\tilde{h}_{b,u}(t)$ is a random variable accounting for small-scale fading.
It is worth noting that the proposed method is based on a model-free DRL framework and is therefore applicable to various antenna configurations and channel models.

We assume that the three networks operate over orthogonal frequency bands with distinct nominal carrier frequencies, such that inter-network interference can be neglected.
This assumption allows us to focus on the variations in service capability across different networks induced by the mobility of satellites and UAVs.
Considering the downlink transmission, the SINR of the $u$-th UAV served by the $b$-th BS is given by
	\vspace{-5pt}
	\begin{equation}\mathrm{\rho}_{u}^{\mathrm{GN}}(t)=\frac{\bar{P}_{b} \big| h_{b,u}(t) \big|^2}{\sum_{j\in\mathcal{B}^{\mathrm{\text{GN}}}\setminus b}\bar{P}_{j}\big| h_{j,u}(t) \big|^2+\sigma_{\mathrm{GN}}^2},\end{equation}
	\vspace{-5pt}
	\begin{equation}\mathrm{\rho}_{u}^{\mathrm{SN}}(t)=\frac{\bar{P}_{b} \big| h_{b,u}(t) \big|^2}{\sum_{j\in\mathcal{B}^{\mathrm{\text{SN}}}\setminus b}\bar{P}_{j}\big| h_{j,u}(t) \big|^2+\sigma_{\mathrm{SN}}^2},\end{equation}
	\vspace{-5pt}
	\begin{equation}\mathrm{\rho}_{u}^{\mathrm{AN}}(t)=\frac{\bar{P}_{b} \big| h_{b,u}(t) \big|^2}{\sum_{j\in\mathcal{B}^{\mathrm{\text{AN}}}\setminus b}\bar{P}_{j}\big| h_{j,u}(t) \big|^2+\sigma_{\mathrm{AN}}^2},\end{equation}
	where  $\sigma^2$ denotes the thermal noise variance at the $u$-th UAV  under different network connections. The achievable rate  \(\mathcal{R}_u(t)\) of the $b$-th BS  (either in GN, SN or AN) to the $u$-th UAV can be computed as
	\vspace{-5pt}
	\begin{equation}
		\mathcal{R}_u(t) = B_u \cdot \mathbb{E}\bigl[\log_2\bigl(1 + \mathrm{\rho}_u(t)\bigr)\bigr],
		\label{eq:sumrate}
	\end{equation}
	where the expectation is taken with respect to the small-scale fading, and $B_u$ denotes the bandwidth allocated to the $u$-th UAV by its serving BS.
Moreover, throughout the flight, each UAV occupies only one physical resource block (PRB) within a given network type, and multiple UAVs are assigned mutually orthogonal time-frequency resources.
	\vspace{-5pt}

	\vspace{-5pt}
	
	\subsection{UAV Trajectory Model}\label{UAV Trajectory Model}

    Without loss of generality, we consider a three-dimensional airspace specified by $[x_L, x_U] \times [y_L, y_U] \times [z_L, z_U]$.
Each UAV flies from a random initial location $\mathbf{q}_u \in \mathbb{R}^{3 \times 1}$ to a common final location $\mathbf{q}_f \in \mathbb{R}^{3 \times 1}$.
This setting applies to scenarios such as parcel collection or aerial inspection \cite{zeng2018cellular}.
Multiple UAVs operate in the same airspace and share the destination $\mathbf{q}_f$, such that their flight experiences can be jointly leveraged to learn coverage-aware navigation policies maintained at the network side.
For the considered SAGIN system, let $T$ denote the mission completion time, and let $\mathbf{q}_u(t) \in \mathbb{R}^{3 \times 1}$, $t \in [0, T]$, denote the position of the $u$-th UAV.
Accordingly, the following holds:
	\vspace{-5pt}
	\begin{align}
		&\mathbf{q}_u(0) = \mathbf{q}_{u}, \quad \mathbf{q}_u(T) = \mathbf{q}_{f},  \\
		\mathbf{q}_L &\preceq \mathbf{q}_u(t) \preceq \mathbf{q}_U, \quad \forall t \in (0, T), 
	\end{align}
	where $\mathbf{q}_L \triangleq [x_L; y_L; z_L]$ and $\mathbf{q}_U \triangleq [x_U; y_U; z_U]$, 
	with $\preceq$ denoting the element-wise inequality.

	\vspace{-5pt}

	\vspace{-5pt}

	\subsection{UAV	Link  Selection Model}
	\label{UAV Handover Model}

In TNs, UAV link selection mainly depends on rapid channel quality variations with propagation distance \cite{Haghrah2023}, whereas in satellite networks, link quality varies slowly due to the high orbital altitude, leading to distinct link selection evaluation criteria \cite{3gpp38821}.
Consequently, when candidate BSs belong to different networks, a unified evaluation framework is required to accommodate the diverse temporal characteristics of their signal dynamics.

Greedily switching to the BS with the highest SINR may lead to an excessively high link switching frequency.
To evaluate the service quality experienced by UAVs during flight, we focus on two key metrics: (i) the link rate $\mathcal{R}_u(t)$, which represents the achievable throughput between the UAV and the serving BS; and (ii) the number of link switches $N_u$, which characterizes the link switching frequency over the mission duration.
Building on this, we introduce a UAV-BS association indicator $b_u(t)$, which denotes the index of the BS serving the $u$-th UAV at time $t$.
A link switching event occurs when $b_u(t)$ changes, and the total number of link switches over the mission duration is defined as
	\vspace{-5pt}
	\begin{equation}
		N_u
		= 
		\sum\nolimits_{t \in (0,T]} 
		\mathbf{1}\!\left\{
		b_u(t^-) \neq b_u(t)
		\right\},
		\label{eq:handover_count}
	\end{equation}
where $b_u(t^-)$ denotes the left-hand limit of $b_u(t)$ at time $t$, and $\mathbf{1}\{\cdot\}$ denotes the indicator function, which equals one if the condition inside the braces holds, and zero otherwise.
This definition captures all discrete link switching events over the continuous-time domain.
Therefore, we jointly employ $\mathcal{R}_u(t)$ and $N_u$ to evaluate the service quality of UAV communications, enabling a unified assessment of candidate BSs.

	\vspace{-5pt}
	\vspace{-5pt}

	\section{Problem Formulation and Transformation}\label{Uav Handover Optimization}

	\subsection{Problem Formulation}\label{Problem Formulation}

In the considered SAGIN system, the objective is to maximize the average link rate of UAVs during flight while minimizing the link switching frequency and flight time, subject to communication QoS constraints.
Accordingly, the optimization problem
\footnote[1]{The switching frequency is characterized by the number of changes in $b_u(t)$ over time.
Since $b_u(t)$ is determined by the received SINR, which is a function of the UAV trajectory $\mathbf{q}(t)$, the switching frequency is implicitly determined by $\mathbf{q}(t)$.}
is formulated as
	\vspace{-5pt}
	\begin{flalign}
		\mathcal{P}_0
        :\; \!
		&	\max_{\{\mathbf{q}(t), b(t)\}}
		\  
		\sum_{u \in \mathcal{U}}
		\{
		\lambda_{1}   \frac{1}{T}\!  \int_{0}^{T}\!
		{
			{\mathcal{R}_u(t)\!}
			\;
		}
		\,dt \!-\! \lambda_{2}  N_u  \!- \!\lambda_{3} T_u 
		\}
		, & \label{eq:11} \\
		{\rm s.t.}\quad &
		b_u(t) \in \{1, \cdots, B\}, \quad \forall u,  \label{eq:9a}  \\
		& \frac{1}{T}\int_{0}^{T}\mathcal{R}_u(t) dt > \mathcal{R}_{\text{req}},  \quad \forall u,
		\label{eq:9b} \\
		& \mathbf{q}_u(0) = \mathbf{q}_u , \quad 
		\mathbf{q}_u(T) = \mathbf{q}_f , \quad \forall u, 
		\label{eq:9c} \\
		& \mathbf{q}_{\mathrm{L}} \preceq \mathbf{q}_u(t) \preceq \mathbf{q}_{\mathrm{U}}, \quad \forall t \in [0,T], \forall u,
		\label{eq:9d} \\
		&
		\vec{\mathbf{v}}_u(t) \cdot \vec{x} > 0 , , \quad  \vec{x} = \mathbf{q}_{uf},\quad \forall t \in [0,T], \forall u,
		\label{eq:9e}  \\
		&
		\|\dot{\mathbf{q}}_u(t)\| \leq V_{\max}, \quad \forall t \in [0, T], \forall u, 
		\label{eq:9f}
	\end{flalign}
where $\lambda_{1}$, $\lambda_{2}$, and $\lambda_{3}$ are weighting coefficients that trade off throughput enhancement, link switching frequency reduction, and flight time minimization, respectively.
$\mathcal{R}_{\mathrm{req}}$ denotes the minimum QoS rate requirement, 
$|\dot{\mathbf{q}}_u(t)|$ represents the instantaneous speed of the $u$-th UAV, 
$V_{\max}$ denotes the maximum UAV speed, 
and $T_u$ denotes the total flight time of the $u$-th UAV.
In $\mathcal{P}_0$, $b_u(t)$ represents the UAV-BS association, while $\mathbf{q}_u(t)$ denotes the UAV position; both jointly determine the SINR and the achievable rate of each link.
Constraint~\eqref{eq:9b} guarantees that each UAV maintains a service rate no lower than the minimum QoS requirement throughout the flight.
Constraints~\eqref{eq:9c} and~\eqref{eq:9d} specify the trajectory boundary conditions, while constraint~\eqref{eq:9e} imposes a directional velocity constraint, ensuring that the UAV continuously moves toward the destination.
	It can be shown that, with the optimal solution to $\mathcal{P}_0$,  UAVs should fly at the maximum speed $V_{\max}$, i.e.,
	$
	\dot{\mathbf{q}}_u(t) = V_{\max} \vec{\mathbf{v}}_u(t),
	\label{eq:vmax}
	$
	where $\vec{\mathbf{v}}_u(t)$ denotes the UAV flying direction vector with $\|\vec{\mathbf{v}}_u(t)\| = 1$.
	Thus, $\mathcal{P}_0$ is expressed as
	\begin{flalign}
		\mathcal{P}_1:\; \!
		&	\max_{\{\mathbf{q}(t), b(t),\vec{\mathbf{v}}(t)\}}
		\  \!
		\sum_{u \in \mathcal{U}}
		\{
		\lambda_{1}   \frac{1}{T}\!  \int_{0}^{T}\!
		{
			\!	{\mathcal{R}_u(t)\!}
			\;
		}
		\,dt \!-\! \lambda_{2}  N_u \!- \!\lambda_{3} T_u
		\}
		, & \label{eq:9} \\
		{\rm s.t.}\quad &
		\text{Constraints } \eqref{eq:9a}\text{--}\eqref{eq:9e}, \\
		&\dot{\mathbf{q}}_u(t) = V_{\max} \vec{\mathbf{v}}_u(t), \quad \forall t \in [0, T], \forall u, \label{eq:9g}  \\
		&
		\|\vec{\mathbf{v}}_u(t)\| = 1, \quad \forall t \in [0, T], \forall u. 
		\label{eq:9h}
	\end{flalign}
	
It can be observed that optimization problem $\mathcal{P}_1$ is a mixed-integer nonlinear programming (MINLP) problem formulated in continuous time, which is generally intractable.
Moreover, in high-mobility scenarios, link switching and trajectory control must be performed in real time.
Hence, due to the inherent non-convexity of the problem and the requirement for continuous-time optimization over the entire mission duration, traditional mathematical methods, such as convex optimization and game theory, are not well suited for this problem.
To address these challenges, we propose a DRL-based approach that enables efficient online decision-making with low inference complexity after offline training.

	\vspace{-5pt}
	
	\vspace{-5pt}
	
	\subsection{Problem Transformation}\label{Problem Transformation}
	
The first step in applying RL algorithms to a real-world problem is to reformulate it as a Markov decision process (MDP).
Since an MDP is defined over discrete time steps, for the joint link selection and trajectory optimization problem $\mathcal{P}_1$, we discretize the time horizon $[0, T]$ into $N$ time steps with a time interval $\Delta t$, where $T = N \Delta t$.
$\Delta t$ is properly chosen such that, within each time step, the large-scale channel gain and the BS antenna gain toward the UAV, i.e., $\beta_{b,u}(\mathbf{q}_u(t))$ and $G_{b,u}(\mathbf{q}_u(t))$ in~\eqref{eq:pb}, remain approximately constant.
Accordingly, the continuous UAV trajectory $\{\mathbf{q}_u(t)\}$ can be approximated by a discrete sequence $\{\mathbf{q}_u(n)\}_{n=1}^{N}$, and the UAV position is given by
	\vspace{-5pt}
	\begin{equation}
		\mathbf{q}_u(n+1) = \mathbf{q}_u(n) + \Delta_s \vec{\mathbf{v}}_u(n), \quad \forall n, 
	\end{equation}
where $\Delta_s = V_{\max} \Delta t$ denotes the UAV displacement per time step, and $\vec{\mathbf{v}}_u(n) \triangleq \vec{\mathbf{v}}(n \Delta t)$ represents the discretized flight direction vector of the $u$-th UAV at the $n$-th time step.
Furthermore, since UAV-BS association is typically determined based on large-scale channel gains to avoid excessively frequent link switching, the associated BS is assumed to remain unchanged within each time step.
Accordingly, the association indicator $b_u(t)$ can be represented in discrete time as $b_u(n)$.
As a result, the received signal power at the $u$-th UAV from the $b$-th BS in~\eqref{eq:pb} can be rewritten as
	\vspace{-5pt}
	\begin{equation}
		\begin{aligned}
			p_{b,u}(n)
			&= \bar{P}_{b}\,\big|h_{b,u}(n)\big|^2 \\
			&= \bar{P}_{b}\,\beta_{b,u}\!\big(\mathbf{q}_u(n)\big)\,
			G_{b,u}\!\big(\mathbf{q}_u(n)\big)\,
			\tilde{h}_{b,u}(n),
		\end{aligned}
		\label{eq:pb_discrete}
	\end{equation}
where $h_{b,u}(n)$ denotes the baseband-equivalent channel between the $b$-th BS and the $u$-th UAV at the $n$-th time step.
The term $p_{b,u}(n)$ represents the received signal power, and $\tilde{h}_{b,u}(n)$ is a random variable accounting for small-scale fading.
At discrete time step $n$, given that the $u$-th UAV is associated with the $b$-th BS, the corresponding SINR is denoted by $\rho_u(n)$, and the achievable rate can be expressed as
\begin{equation}
\mathcal{R}_u(n) = B_u \cdot \,\mathbb{E}\!\left[\log_2\!\big(1+\rho_u(n)\big)\right].
\end{equation}
Accordingly, the total number of link switches defined in~\eqref{eq:handover_count} can be approximated as
	\vspace{-5pt}
	\begin{equation}
		N_{u} \triangleq \sum\nolimits_{n=2}^{N} \mathbf{1}\!\left\{ b_u(n) \neq b_u(n-1) \right\}.
		\label{eq:ho_count_discrete}
	\end{equation}
	
	Based on the above discussion, $\mathcal{P}_1$ can be approximated as
	\vspace{-5pt}
	\vspace{-5pt}
	\begin{flalign}
		\mathcal{P}_2:\; \!
		&	\max_{\{\mathbf{q}(n),\vec{\mathbf{v}}(n),b(n)\}} 
		\sum_{u \in \mathcal{U}}
		\{
		\lambda_{1}\, \!\frac{1}{N}\!\sum_{n=1}^{N}\! {\mathcal{R}_u(n)\!}\! \;\!-\!\;\!\! \lambda_{2}\, N_{u} - \!\!\lambda_{3}  T_u
		\}	,
		\label{eq:P1_obj}\\[2pt]
		{\rm s.t.}\quad &
		b_u(n) \in \{1,\ldots,B\}, \quad \forall n, u,
		\label{eq:P1_c1}\\
		& \mathbf{q}_u(n+1)=\mathbf{q}_u(n)+\Delta_s\,\vec{\mathbf{v}}_u(n),\quad \forall n, u,
		\label{eq:P1_c3}\\
		& \frac{1}{N}\sum_{n=1}^{N} \mathcal{R}_u(n) \;\ge\; \mathcal{R}_{\mathrm{req}}, \forall u,
		\label{eq:9ccc} \\
		& \mathbf{q}_L \preceq \mathbf{q}_u(n) \preceq \mathbf{q}_U,\quad \forall n,  u, 
		\label{eq:P1_c4}\\
		&
		\mathbf{q}(0)=\mathbf{q}_u,\qquad \mathbf{q}(N)=\mathbf{q}_f,  u,
		\label{eq:P1_c5}\\
		&
		\vec{\mathbf{v}}_u(n) \cdot \vec{x} > 0  , \quad  \vec{x} = \mathbf{q}_{uf}, \quad \forall n, u, \label{eq:P1_c6}
		\\
		&
		\|\vec{\mathbf{v}}_u(n)\| = 1, \quad \forall n, u.
		\label{eq:P1_c7}
	\end{flalign}

    It can be observed that problem $\mathcal{P}_2$ involves a hybrid action space with two types of decision variables: the discrete UAV-BS association indicator $b_u(n)$ and the continuous UAV trajectory  $\mathbf{q}(n)$.
Moreover, these two decision variables are strongly coupled and highly intertwined.
Such a large hybrid action space significantly increases the decision-making complexity and may destabilize the learning process, rendering conventional DRL methods not well suited for this joint decision-making task.

	\vspace{-5pt}

	\section{Proposed HDRL Algorithm}\label{Proposed HDRL Algorithm}

	\subsection{HDRL Framework}
	\label{HDRL Framework}
	
To address the challenges arising from the large hybrid action space and the strong coupling between discrete and continuous decision variables, we propose an HDRL framework composed of hierarchically organized DRL modules operating at different time scales.
The proposed framework operates over two hierarchical levels: (i) a top-level module, described in Section~\ref{sec:ddqn}, which takes the system state as input and determines the UAV-BS association; and (ii) a lower-level module, presented in Section~\ref{sec:SAC}, which leverages both the system state and the selected BS association to generate continuous UAV trajectory  actions until the current association terminates.
Once the association expires, the top-level module selects a new BS association, and the above two-level decision-making process is repeated.
The framework is trained using gradient-based optimization at different temporal scales, where the lower-level and top-level modules are optimized with respect to intrinsic rewards (Section~\ref{sec:SAC}) and extrinsic rewards (Section~\ref{sec:ddqn}), respectively.

	\begin{figure*}[t]
		\centering
		\includegraphics[width=6.5in]{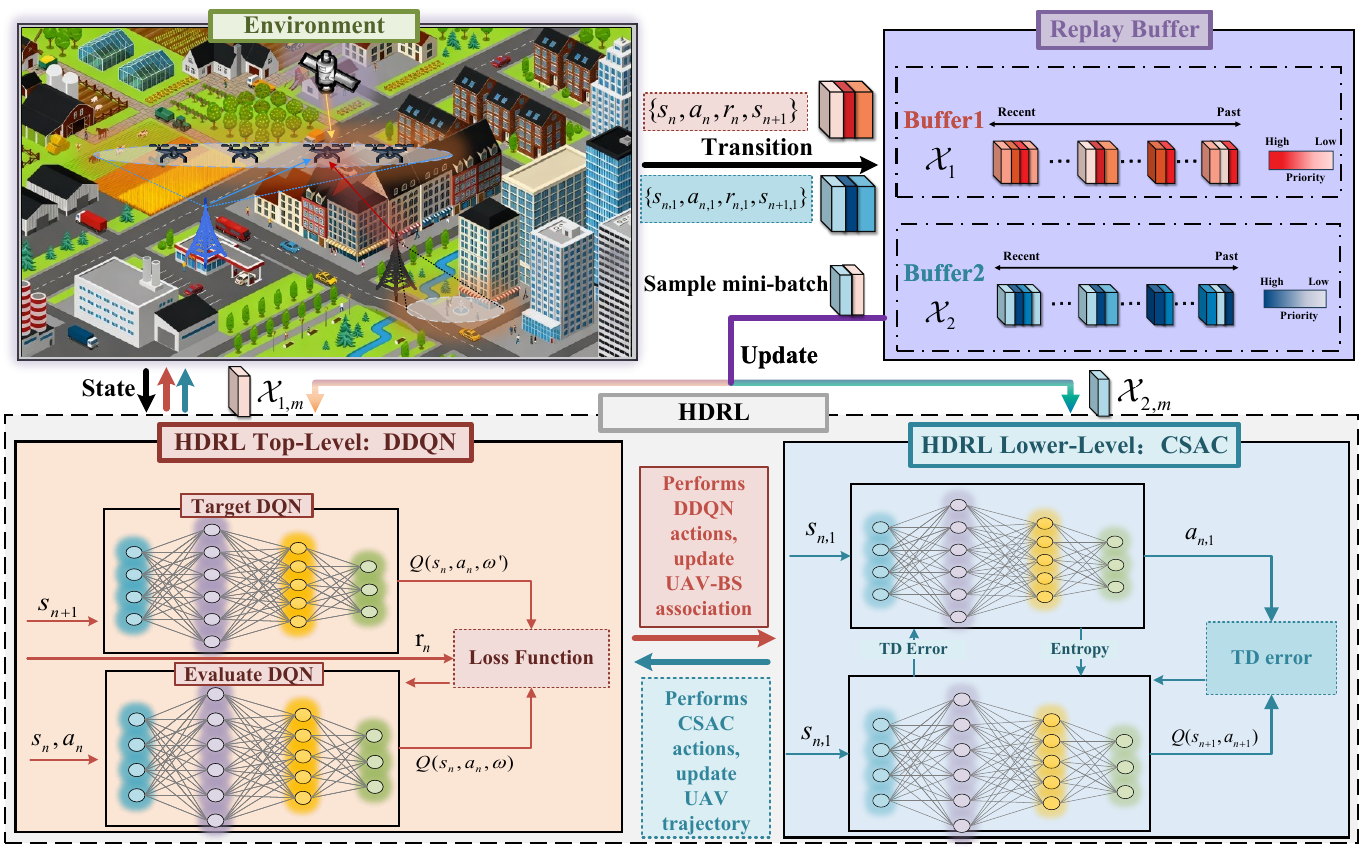}
		\caption{The framework of the proposed HDRL algorithm for solving joint optimization of link selection and trajectory control.}
		\label{fig2}
	\end{figure*}

As illustrated in Fig.~\ref{fig2}, the proposed HDRL framework consists of three main components: the environment, the replay buffer, and the HDRL network.
The environment provides the UAVs with the current system state and returns the corresponding reward signal after each executed action.
The HDRL network comprises a top-level DDQN, which outputs UAV-BS association decisions, and a lower-level CSAC, which generates continuous trajectory control actions; the resulting interaction data are then fed back to the environment.
The replay buffer categorizes and stores the transition samples generated by the two hierarchical levels and supplies them for policy updates, thereby forming a closed-loop learning process.
Details of the DDQN and CSAC designs are provided in Sections~\ref{sec:ddqn} and~\ref{sec:SAC}, respectively, while the overall HDRL algorithm is presented in Section~\ref{Overall HDRL Algorithm}.

	\vspace{-2pt}

	\vspace{-5pt}
	
	\subsection{HDRL Top-Level: DDQN for UAV-BS Association}\label{sec:ddqn}

Unlike terrestrial networks, the heterogeneous architecture of SAGIN, characterized by diverse coverage patterns, renders UAV-BS association decisions highly complex, making the direct application of DRL difficult to train and prone to slow convergence.
To address this issue, we design a DDQN-based top-level agent that maps the hybrid association space into only two semantic actions: \textit{Remain} and \textit{Switch}.
By leveraging double Q-value estimation to decouple action selection from action evaluation, the top-level agent effectively suppresses overestimation bias and achieves stable and fast-converging UAV-BS association policies.
The state space, action space, reward function, and the DDQN algorithm are defined sequentially as follows.

	\vspace{-5pt}
	
	\subsubsection{Top-Level State Space}
	
	\vspace{0.5em}

The state of the UAV at the $n$-th time step, denoted by $\mathbf{s}_n$, consists of three components.
The first component is the UAV-BS association indicator $b_n$, where $b_n = b \in \mathcal{B}$ denotes the index of the serving BS.
The second component is the link rate between the UAV and its associated BS at the $n$-th time step, denoted by $R_n$.
The third component is the UAV position $\mathbf{q}_n = (x_n, y_n, z_n)$.
Accordingly, the state $\mathbf{s}_n$ is given by
	\vspace{-5pt}
	\begin{equation}
		\mathbf{s}_n \triangleq \big[\,b_{n},\; R_{n},\; \mathbf{q}_n \big]^{T}.
	\end{equation}
	
	\vspace{-5pt}
	
	\subsubsection{Top-Level Action Space}
	To avoid action space expansion, non-stationarity, and unstable training caused by exhaustively considering all BSs, we model each UAV as an agent with two types of actions: \textit{Remain}, which maintains the current association, and \textit{Switch}, which updates the association to the candidate BS with the highest state-action value (Q-value).
Specifically, the action $a_n \in \mathcal{A}$ can be denoted by
	\begin{equation}
		a_n = \{\textit{Remain},\, \textit{Switch}\},
	\end{equation}
	After executing action~$a_n$, the UAV-BS association indicator is updated from the previous connection state $b_n$ to $b_{n+1}$.
If $b_{n+1} = b_n$, the UAV maintains its connection to the same BS; otherwise, the UAV switches to a new BS.
In the top-level decision process, the designed state space retains only the three most relevant elements for UAV-BS association, thereby achieving a minimally sufficient representation for link selection.
Meanwhile, the original action space over numerous candidate BSs is compressed into a binary set, which substantially reduces the complexity of the top-level decision-making.

	\subsubsection{Top-Level Reward Function}

	To encourage the UAV to maintain a high link rate throughout its mission \eqref{eq:P1_obj}, the rate reward $r_{n}^R$ is defined as
	\vspace{-5pt}
	\begin{equation}
		r_{n}^R = 
		({R_{n+1} - \min\nolimits_{E}\{R\}})/(\max\nolimits_{E}\{R\} - \min\nolimits_{E}\{R\}),
	\end{equation}
	where $R_{n+1}$ denotes the  link rate of UAV after executing action~$a_n$ and transitioning to the next state~$s_{n+1}$. 
	The terms $\max_{E}\{R\}$ and $\min_{E}\{R\}$ represent the maximum and minimum link rates, respectively.
	To reduce the  link switching frequency in \eqref{eq:P1_obj}, a  penalty $r_{n}^M$ is introduced as
	\vspace{-5pt}
	\begin{equation}
		r_{n}^M =
		\begin{cases}
			0, & \text{if } b_{n+1} = b_{n}, \\[4pt]
			-1, & \text{if } b_{n+1} \neq b_{n}.
		\end{cases}
	\end{equation}
	Therefore, the overall extrinsic reward  consists of two components: a rate reward and a link switching  penalty, which can be expressed as
	\vspace{-5pt}
	\begin{equation}
		r_n(s_n,a_n) 
		= 	\lambda_{1} \cdot \, r_n^{R} + 	\lambda_{2} \cdot \, r_n^{M},
	\end{equation}
	where $	\lambda_{1}$ and $	\lambda_{2}$ are weighting factors that balance the
	rate reward and association-switching penalty  in \eqref{eq:P1_obj}.

	\subsubsection{DDQN Algorithm}\label{DDQN}
The UAV-BS association problem can be formulated as the maximization of the expected discounted cumulative reward over the time horizon, which is defined as
$
G = \sum_{n=1}^{N} \gamma_1^{\,n-1} r_n(s_n, a_n),
$
where $0 \le \gamma_1 \le 1$ is a discount factor.
The objective of the agent is to obtain the optimal decision policy $\pi^{\ast}$ (i.e., selecting the optimal UAV-BS association $b(n)$) that maximizes $G$, which can be expressed as $\pi^{\ast} = \arg\max_{\pi} G$.
To handle the time-varying candidate BS set and mitigate the ping-pong switching effects caused by satellite and UAV mobility, we employ a DDQN-based agent to learn the optimal policy $\pi^{\ast}$.
In the DDQN framework, the optimal policy can be expressed as
$
\pi^{\ast} = \arg\max_{a} Q^{\pi^{\ast}}\!\big(s_n, a_n\big),
$
where $Q^{\pi^{\ast}}$ denotes the state-action value function, which is defined as
$
Q^{\ast}(s_n, a_n)
= \mathbb{E}_{\pi}\!\left[
\sum_{k=0}^{\infty} \gamma_1^{k} r_{n+k+1}
\;\middle|\;
s_n = s,\; a_n = a
\right].
$

	At each decision step $n$, a deep neural network (DNN) parameterized by $\bm{\omega}$ is employed to approximate the value function, while advanced network modules are also applicable \cite{wang2025tensor}. 
	Specifically, the evaluation network $Q(s,a;\bm{\omega})$  takes the current state $s_n$ as input and outputs the Q-values corresponding to all possible actions $a \in \mathcal{A}$, 
	which serve as the basis for action selection.
	To balance the trade-off between {exploration} and {exploitation}, 	we then adopt the $\varepsilon$-greedy policy
	\vspace{-5pt}
	\begin{equation}
		a_n=
		\begin{cases}
			\arg\max\limits_{a\in\mathcal{A}} Q(s_n,a;\bm{\omega}), & \text{with prob. } 1-\varepsilon, \\[6pt]
			\text{randomly select } a \in \mathcal{A}, & \text{with prob. } \varepsilon,
		\end{cases}
		\label{eq:eps_greedy}
	\end{equation}
	where $\varepsilon$ exponentially decays from $\varepsilon_{\text{init}}$ to $\varepsilon_{\text{final}}$, 
	and remains fixed at $\varepsilon_{\text{final}}$ thereafter. 
	After executing action $a_n$, the agent receives an immediate reward $r_n $ and transitions to the next state $s_{n+1}$, where  $\{s_n, a_n, r_n, s_{n+1}\}$ is stored in the replay buffer $\mathcal{X}_1$, which can be seen in Fig.  \ref{fig2}. 
	During the network updating stage, a mini-batch of experiences $\mathcal{X}_{1,z}$ is randomly sampled from $\mathcal{X}_1$. 
	For each sampled experience, the target value is defined as the sum of the immediate reward and the discounted maximum Q-value of the next state, given by
	\vspace{-5pt}
	\begin{equation}
		y_n = r_n + \gamma_1 Q(s_{n+1}, \hat{a}_{n+1}; \bm{\omega}'),
		\label{the target value}
	\end{equation}
	where $\hat{a}_{n+1} = \arg\max_{a \in \mathcal{A}} Q(s_{n+1}, a; \bm{\omega})$, 
	and $\bm{\omega}'$ denotes the parameters of the target network. 
	The target network is employed solely for computing the target value $y_n$ 
	and does not participate in backpropagation to ensure training stability.
	The evaluation network is optimized by minimizing the mean square error (MSE) loss function
	\vspace{-5pt}
	\begin{equation}
		L(\bm{\omega}) = 
		\mathbb{E}_{\{s,a,r,s_{n+1}\} \in \mathcal{X}_{1,z}}
		\Bigl[\bigl(y_n - Q(s_n, a_n; \bm{\omega})\bigr)^2\Bigr].
		\label{loss function}
	\end{equation}
	The target network, evaluation network, and the loss function of the DDQN network can be observed in the lower-left part of Fig.~\ref{fig2}.	
	The network parameters are optimized using the Adam optimizer, given by \cite{kingma2014adam}
	\vspace{-5pt}
	\begin{equation}
		\bm{\omega} \leftarrow \text{Adam}\!\left(\nabla_{\bm{\omega}} L(\bm{\omega}),\, \eta_1\right),
		\label{eq:update}
	\end{equation}
	where $\eta_1$ is the learning rate.
	To further enhance the training stability, the parameters of the training network are periodically copied to the target network 
	every $\lambda_U$ iterations by
	$
	\bm{\omega}' \leftarrow \bm{\omega}
	$.
	Unlike the basic DQN, our DDQN framework incorporates an adaptive exploration strategy, experience replay, and a dual-network architecture in Fig.~\ref{fig2}, which not only alleviates Q-value overestimation and training oscillations in non-stationary environments but also significantly enhances the stability, convergence efficiency, and robustness of policy learning in highly dynamic SAGIN scenarios.
	
	\vspace{-5pt}
	\vspace{-5pt}
	
	\subsection{HDRL Lower-Level:  CSAC  for  UAV Trajectory Control}\label{sec:SAC}

Existing studies commonly discretize the continuous action space of UAV trajectory control, which not only introduces throughput loss but also depends on extensive heuristic rules for designing complex reward-penalty structures and numerous hyperparameters, thereby limiting their applicability in multi-constraint environments.
Building on this, we propose a Lagrangian-based CSAC algorithm to optimize UAV trajectory in a continuous action space. The algorithm not only eliminates cumbersome reward and hyperparameter engineering but also naturally captures the nonlinear couplings between states and decision variables under multiple constraints.

	\subsubsection{Lower-Level State Space}
	As illustrated in Fig.~\ref{fig2}, the top-level first generates and holds the UAV-BS association decision, upon which the lower-level produces the trajectory control actions, and the resulting new state is then fed back to the top-level to enable the update of the subsequent  association decision in Section~\ref{HDRL Framework}.
	For the  $1$-th lower-level time step corresponding to the $m$-th top-level time step,
	the UAV state $\mathbf{s}_{m,1}$ consists of three components:
	(i) the UAV-BS association indicator  $b_{m}$, 
	(ii) the link rate
	$R_{m,1}$, 
	and (iii) the UAV position vector $\mathbf{q}_{m,1} = (x_{m,1}, y_{m,1}, z_{m,1})$. 
	Thus, the state of the  $1$-th lower-level time step corresponding to the $m$-th top-level time step can be expressed as
	\vspace{-5pt}
	\begin{equation}
		\mathbf{s}_{m,1} \triangleq 
		[\, b_{m}, \; R_{m,1}, \; \mathbf{q}_{m,1} \,]^{T}.
		\label{eq:state}
	\end{equation}
	
	\subsubsection{Lower-Level Action Space}
	
	The action space of the agent is continuous, and each action  represents the control of the UAV’s trajectory direction at a time step. 
	For each UAV, its new position in the next step is obtained by adding the current position and the action-induced displacement. 
	The action $a_{m,1} = \vec{\mathbf{v}}_{m,1}$ corresponds to the continuous selection of the UAV’s velocity control at each time step,
	where the velocity vector is treated as a continuous control variable in both direction and magnitude. 
	Thus, the new UAV position $\mathbf{q}_{m,2}$ after taking the action can be expressed as
	\vspace{-5pt}
	\begin{equation}
		\mathbf{q}_{m,2} = \mathbf{q}_{m,1} + a_{m,1} V_{\max} \Delta t,
		\label{eq:action}
	\end{equation}

	\subsubsection{Lower-Level Reward Function}
	
	To guide the UAV to fly toward regions with better communication quality \eqref{eq:P1_obj}, the  rate reward
	$r_{m,1}^{\mathrm{rate}}$ is defined as  
	\vspace{-5pt}
	\begin{equation}
		r_{m,1}^{\mathrm{rate}}\left({s}_{m,1}, a_{m,1}\right)
		= \frac{R_{m,2} - \min\nolimits_{E}\{R\}}
		{\max\nolimits_{E}\{R\} - \min\nolimits_{E}\{R\}},
		\label{eq:reward}
	\end{equation}
	where $R_{m,2}$ denotes the  link rate  after taking  action $a_{m,1}$ and reaching the next state ${s}_{m,2}$.
	To encourage the agent to minimize the flight time \eqref{eq:P1_obj}, a goal-approaching reward $r_{m,1}^{\mathrm{goal}} $is defined as 
	\vspace{-5pt}
	\begin{equation}
		r_{m,1}^{\mathrm{goal}}
		=
		\!\left(
		{D_{\max} - d_{m,2}}
		\right)/(D_{\max} - D_{\min}),
		\label{eq:goal_reward}
	\end{equation}
	where  $D_{\min}$ and $D_{\max}$ denotes the initial distance between the UAV and the destination at the beginning of the mission and $d_{m,2}$ is the Euclidean distance from the UAV's current position to the target.
	Building on this, the future intrinsic reward function is formulated as 
	\vspace{-5pt}
	\begin{equation} r_{m,1}(s_{m,1},a_{m,1}) = \lambda_{1} \cdot r_{m,1}^{\mathrm{rate}} + \lambda_{3} \cdot r_{m,1}^{\mathrm{goal}}, 
		\label{eq:total_reward} 
	\end{equation} 
	where $\lambda_{1}$ and $\lambda_{3}$ denote the  weighting factors that balance the communication quality and flight-efficiency objectives in \eqref{eq:P1_obj}.

	\subsubsection{Lower-Level  Cost Function}
	To ensure that the UAV satisfies both trajectory and QoS constraints during flight \eqref{eq:P1_c3}\eqref{eq:9ccc}\eqref{eq:P1_c4}\eqref{eq:P1_c5}, a cost function is introduced.
	To guide the UAV to fly toward regions with more stable and higher-quality communication links \eqref{eq:9ccc}, the minimum rate requirement $\mathcal{R}_{\mathrm{req}}$ represents the threshold of acceptable communication performance.
	Specifically,  the QoS cost $c^{\mathrm{qos}}_{m,1}$
	is defined as 
	\vspace{-5pt}
	\begin{equation}
		c^{\mathrm{qos}}_{m,1}
		=
		\,
		\!\left(
		{R_{m,2} -R_{\mathrm{req}}}
		\right)/R_{\mathrm{req}}.
		\label{eq:qos_reward}
	\end{equation}
	When  $R_{m,2} \ge R_{\mathrm{req}}$, the QoS cost becomes non-negative, representing a reward; otherwise, the QoS cost becomes a penalty.
	To constrain the UAV within the designated operational region in \eqref{eq:P1_c3}\eqref{eq:P1_c4}\eqref{eq:P1_c5},
	a boundary penalty is imposed whenever the UAV’s position violates this constraint
	\vspace{-5pt}
	\begin{equation}
		c^{\mathrm{bnd}}_{m,1}
		=
		-\eta_{\mathrm{bnd}}\,
		\mathbb{I}\!\left(
		\mathbf{q}_{m,2} \notin [\,\mathbf{q}_{\mathrm{L}},\, \mathbf{q}_{\mathrm{U}}\,]
		\right),
		\label{eq:boundary_reward}
	\end{equation}
	where $\eta_{\mathrm{bnd}}>0$ is the boundary-penalty weight and $\mathbb{I}(\cdot)$ is an indicator function that equals one if the condition holds and zero otherwise.
	When the UAV remains within the feasible region, $c^{\mathrm{bnd}}_{m,1}=0$;
	otherwise, a negative penalty is applied.  
	By designing cost functions that are in one-to-one correspondence with the constraints in problem $\mathcal{P}_2$,
	we explicitly recast the original hard constraints as learnable cost signals, enabling the CSAC agent to adaptively balance  optimization of the task objective and constraints during training.
	
	\subsubsection{CSAC Algorithm} 
	
	In the lower-level of the HDRL framework, our objective is to derive, for each UAV, an optimal continuous control policy that maximizes its long-term cumulative return over the entire flight period while satisfying trajectory and QoS constraints.
	Specifically, the optimal policy can be formulated as 
	\vspace{-5pt}
	\begin{align}
		\max_{\pi} \;
		&\mathbb{E}_{\pi}\!\left[
		\sum\nolimits_{n=0}^{\infty} \gamma_2^{n} r(s_{n}, a_{n})
		\right]
		\nonumber\\
		\text{s.t.} \quad
		&\mathbb{E}_{\pi}\!\left[
		\sum\nolimits_{n=0}^{\infty} \gamma_2^{n} c_{k}(s_{n}, a_{n})
		\right]
		\le d_{k}, \; \forall k,
		\label{eq:constraint_rl}
	\end{align}
	where $r(s_{n}, a_{n})$ denotes the instantaneous reward, 
	$c_{k}(s_{n}, a_{n})$ represents the cost function associated with the $k$-th constraint, 
	and $d_{k}$ is the corresponding tolerance threshold.
To solve the above constrained optimization problem, we adopt the CSAC algorithm, which introduces a Lagrangian multiplier to explicitly embed the constraint terms into the objective function. 
This design enables the agent to maximize the expected cumulative reward while ensuring constraint satisfaction throughout the training process.

	Specifically, we first define the soft state-action value function 
	based on the policy $\pi$ as 
	\vspace{-5pt}
	\begin{align}
		&Q^{\pi}(s_{n}, a_{n}) = \nonumber\\
		&\mathbb{E}_{\pi}\!\Bigg[
		\sum_{n}
		\Big(
		r(s_n,a_n)
		- \sum_{i=1}^{K}\lambda_i\,c_i(s_n,a_n)
		+ \alpha\,\mathcal{H}\!\big(\pi(\cdot|s_n)\big)
		\Big)
		\Bigg],
		\label{eq:csac_lagrange}
	\end{align}
	where $\lambda_{i} \ge 0$ denotes the Lagrange multiplier corresponding to the $i$-th constraint, 
	$\alpha$ is the temperature coefficient, and 
	$\mathcal{H}\!\big(\pi(\cdot\!\mid s_n)\big)$ is the policy entropy term for actions at state~$s_n$, 
	where $\pi(\cdot\!\mid s_n)$ denotes the distribution of the action selection.
	$\mathcal{H}\!\big(\pi(\cdot\!\mid s_n)\big)$ can be defined as
	$
	\mathcal{H}\!\big(\pi(\cdot\!\mid s)\big)
	= -\mathbb{E}_{a\sim\pi(\cdot\mid s)}\!\big[\log \pi(a\!\mid s)\big].
	$
	By adopting the policy entropy term, SAC makes the probability
	distribution of continuous actions more uniform, leading
	to improved generalization capability and exploration ability.
	
	The SAC framework consists of one actor network $\pi(\phi)$, which outputs a
	probabilistic action distribution, and four Q-critic networks that estimate the
	state-action values, including two evaluation networks
	$Q(\theta_1)$ and $Q(\theta_2)$ and their corresponding target networks
	$Q(\theta'_1)$ and $Q(\theta'_2)$ in  Fig.~\ref{fig2}. At initialization,
	$\theta'_1\!\leftarrow\!\theta_1$ and $\theta'_2\!\leftarrow\!\theta_2$.
	At each step, for each agent, the actor network $\pi(\phi)$ takes the state~$s_n$ as input and outputs the action distribution, 
	including a mean scalar $\mu(s_n;\phi)$ and a standard deviation (STD) logarithm scalar $\hat{\sigma}(s_n;\phi)$. 
	To control the range of STD, we clip $\hat{\sigma}(s_n;\phi)$ into $[-20,\,2]$ and get $\hat{\sigma}_{\mathrm{clip}}(s_n;\phi)$, 
	then the STD can be obtained by
	$
	\sigma(s_n;\phi) = \exp\!\big(\hat{\sigma}_{\mathrm{clip}}(s_n;\phi)\big).
	$
	SAC applies the reparameterization trick to sample the action, thereby reducing the variance of gradient estimation 
	during back propagation~\cite{haarnoja2018soft}:
	\vspace{-5pt}
	\begin{equation}
		a = \Delta P \cdot \tanh\!\big[\mu(s_n;\phi) + \sigma(s_n;\phi)\cdot\xi\big],
		\label{eq:action_sample}
	\end{equation}
	where $\xi \sim \mathcal{N}(0,1)$ is a Gaussian random number. 
	The scaling coefficient~$\Delta P$ and the $\tanh(\cdot)$ function are used to map the action value to the desired range. 
	After executing the action~$a_n$, the lower-level agent receives the reward~$r_n$ and next state~$s_{n+1}$, 
	and this experience $\{s_n,a_n,r_n,s_{n+1}\}$ is stored in the replay memory~$\mathcal{X}_2$. 
	Subsequently, similar to the DDQN algorithm, SAC randomly samples a mini-batch of experience 
	$\mathcal{X}_{2,z}$ from $\mathcal{X}_2$ to train the  networks in Fig.~\ref{fig2}.

	The Q-critic networks are expected to estimate more accurate state-action values, so each Q-critic network 
	(for $i=1,2$) can be trained by minimizing the loss function denoted as
	\vspace{-5pt}
	\begin{equation}
		L(\theta_i) =
		\mathbb{E}_{\mathcal{X}_{2,z}}
		\Big[
		(Q(s_n,a_n;\theta_i) - y(r_n,s_{n+1}))^2
		\Big],
		\label{eq:critic_loss}
	\end{equation}
	where the estimation target soft-Q value $y(r_n,s_{n+1})$ is given by
	$
	y(r_n,s_{n+1})\! \!=\!\!
	r_n \!\!+ \!\gamma_2
	\Big(
	\min_{i=1,2} Q(s_{n+1},a_{n+1};\theta'_i)
	- \alpha\log\pi(a_{n+1}\!\mid\!s_{n+1};\phi)
	\Big)
	\label{eq:softQ}
	$,
	where $a_{n+1}$ is the next action sampled from the actor network with input~$s_{n+1}$, 
	and $\pi(a_{n+1}\!\mid\!s_{n+1};\phi)$ is the probability of~$a_{n+1}$ in the action distribution. 
	Based on the gradients of~\eqref{eq:critic_loss}, the parameters of the two evaluation Q-critic networks $\theta_1$ and $\theta_2$ 
	can be updated similar to~\eqref{eq:update}. 
	Then, the parameters of the two target Q-critic networks $\theta'_1$ and $\theta'_2$ 
	can be softly updated by
	\vspace{-5pt}
	\begin{equation}
		\theta'_i \leftarrow \tau \theta_i + (1 - \tau)\theta'_i, 
		\qquad i \in \{1,2\},
		\label{eq:soft_update}
	\end{equation}
	where $0 < \tau \ll 1$ is the soft updating factor.
	The actor network is expected to output actions with a larger composite metric including both Q-value and policy entropy. 
	It can be trained by minimizing the loss function given by
	\vspace{-5pt}
	\begin{equation}
		L(\phi) =
		\mathbb{E}_{\{s_{n}\}\in\mathcal{X}_{2,m}}
		\Big[
		\alpha \log\pi(\tilde{a}_{n}\!\mid s_{n};\phi)
		- \min_{i=1,2} Q(s_{n},\tilde{a}_{n};\theta_i)
		\Big],
		\label{eq:actor_loss}
	\end{equation}
	where $\tilde{a}_{n}=\pi(s_{n};\phi)$ is the action sampled from the current actor network with input~$s_{n}$. 
	Based on the gradients of~\eqref{eq:actor_loss}, the parameters of the actor network~$\phi$ can be updated similar to~\eqref{eq:update}.
	The temperature~$\alpha$ can be automatically updated by minimizing the loss function in Fig.~\ref{fig2}:
	\vspace{-5pt}
	\begin{equation}
		L(\alpha) =
		\mathbb{E}_{\{s_{n}\}\in\mathcal{X}_{2,m}}
		\Big[
		-\alpha
		\big(
		\mathcal{H}_{\mathrm{target}} + \log\pi(\tilde{a}_{n}\!\mid s_{n};\phi)
		\big)
		\Big],
		\label{eq:alpha_loss}
	\end{equation}
	where the target entropy is set as $\mathcal{H}_{\mathrm{target}}=-1$, i.e., the negative value of the action dimension.
	During each training iteration, 
	in addition to updating the actor and critic networks, 
	CSAC adaptively adjusts the Lagrange multipliers $\lambda_i$ for each constraint according to
	\vspace{-5pt}
	\begin{equation}
		\lambda_i
		\leftarrow
		\Big[
		\lambda_i
		+ \eta_{\lambda}
		\big(\mathbb{E}[c_i(s_{n},a_{n})] - d_i\big)
		\Big]_+,
		\label{eq:lambda_update_csac}
	\end{equation}
	where $\eta_{\lambda}$ is the learning rate for the multipliers, 
	and $[\cdot]_+$ denotes projection onto the nonnegative space. 
	When the expected constraint cost $\mathbb{E}[c_i(s_{n},a_{n})]$ exceeds its limit $d_i$, 
	the corresponding $\lambda_i$ increases to impose a stronger penalty; 
	otherwise, $\lambda_i$ decreases gradually. 
	In summary, the proposed CSAC algorithm explicitly incorporates Lagrangian multipliers to enable joint learning of continuous trajectory control and multiple constraints in \eqref{eq:constraint_rl}, which maintains high exploration capability and training stability, 
while effectively avoiding cumbersome reward and hyperparameter engineering.

	\begin{algorithm}[t]
		\caption{HDRL Algorithm}
		\label{alg:HDRL}
		\begin{algorithmic}[1]
			\Require  $\varepsilon$, $\tau$,  $\gamma_1$, $\gamma_2$, $\eta_1$, $\eta_2$, $\eta_\lambda$.
			\State \textbf{Initialization:} Set $t = 0$ and initialize 
			$\mathcal{X}_1,Q(\bm{\omega}),Q(\bm{\omega}')$
			of DDQN agent and  $\mathcal{X}_2$, $\pi(\phi)$,$Q(\theta_1), Q(\theta_2)$ of CSAC agent.
			\State  Randomly initialize weights  $\left( \bm{\omega},\bm{\omega}' \right)$ and  $\left(\theta_1, \theta_2, \alpha, \{\lambda_i\}  \right)$.
			\For{each UAV}
			\For{episode $=$ $1$ to $M$}
			\For{each time step $n=0,1,\ldots,N$}
			\State Observe  $s_n$ and 
			execute  $a_n$ in \eqref{eq:eps_greedy}.
			\State Observe $r_n$  and the next state ${s}_{n+1}$.
			\State Store  $({s}_n,a_n,r_n,{s}_{n+1})$ into $\mathcal{X}_1$.
            \State Randomly sample a mini-batch $\mathcal{X}_{1,m}$ from $\mathcal{X}_1$.
			\For{each sample $ \in \mathcal{X}_{1,m}$}
			\State Compute the target value in \eqref{the target value}.
            \State Update network parameters by \eqref{loss function} \eqref{eq:update}.
			\EndFor
			\If{$n \bmod \lambda_U = 0$}
			\State Update target network parameters: $\bm{\omega}' \leftarrow \bm{\omega}$
			\EndIf
			\State  Observe $s_{n,1}$ and sample action $a_{n,1}$ by \eqref{eq:action_sample}.
			\State Observe  ${s}_{n,2}$, $r_{n,1}$, and $c_{n,1}$.
			\State Store   $({s}_{n,1},a_{n,1},r_{n,1},{s}_{n,2},c_{n,1})$  into  $\mathcal{X}_2$.
     		\State Randomly sample a mini-batch $\mathcal{X}_{2,m}$ from $\mathcal{X}_2$.
			\For{each sample $\in\!\mathcal{X}_{2,m}$}
			\State Compute the target value by \eqref{eq:csac_lagrange}.
            \State Update  critics by \eqref{eq:critic_loss} and actor by \eqref{eq:actor_loss}.
			\State Update the temperature by \eqref{eq:alpha_loss}. 
            \State  Update the multipliers by \eqref{eq:lambda_update_csac}.
            \State Update the  target networks by \eqref{eq:soft_update}.
			\EndFor
			
			\State  Update the top-level state by $s_{n+1}\!\leftarrow\!{s}_{n,2}$.
			\EndFor
			\EndFor
			\EndFor
			\Ensure 
			$
			\pi_{\mathrm{T}}^{*}(a|\mathbf{s};\bm{\omega}^{*})$,
			$
			\pi_{\mathrm{L}}^{*}(a|\mathbf{s};\phi^{*})$,
			$
			Q_{\mathrm{T}}^{*}(\bm{\omega}^{*})$,
			$
			Q_{\mathrm{L}}^{*}(\theta_1)$,
			$
			Q_{\mathrm{L}}^{*}(\theta_2)$,
			$
			\alpha^{*}$,
			$
			\{\lambda_i^{*}\}
			$.
		\end{algorithmic}
	\end{algorithm}

	\vspace{-4mm}
	\subsection{Overall HDRL Algorithm}\label{Overall HDRL Algorithm}

The detailed pseudo-code of the proposed algorithm is provided in Algorithm~\ref{alg:HDRL} and explained as follows.
Steps~1--2 initialize the network parameters and the initial states.
Each training episode starts from the initial states of all UAVs.
In Steps~4--13, the top-level DDQN agent in Section~\ref{sec:ddqn} collects extrinsic transition samples and updates the top-level network parameters.
In Steps~14--24, the lower-level CSAC agent in Section~\ref{sec:SAC} executes continuous trajectory control conditioned on the top-level decisions, collects intrinsic transition samples, and updates the lower-level network parameters.
Finally, the top-level DDQN agent learns the optimal UAV-BS association policy, while the lower-level CSAC agent converges to the optimal UAV trajectory control policy, as illustrated in Fig.~\ref{fig2}.
By hierarchically structuring the decision-making process and optimizing each level at its own temporal scale in Section~\ref{HDRL Framework}, the top-level and lower-level decisions are effectively decoupled and learned at different temporal scales, which naturally reduces the dimensionality of the hybrid action space, lowers the computational complexity, and improves training stability.

	\begin{table}[t]
		\centering
		\footnotesize   
		
		\captionsetup{font=scriptsize}
		
		\caption{Main System-Level Experiment Parameters}
		\label{tab:main_params}
		\setlength{\tabcolsep}{6pt}
		\begin{tabular}{p{0.55\columnwidth} p{0.37\columnwidth}}
			\toprule
			\textbf{Parameter} & \textbf{Value} \\
			\midrule
			\textbf{GN parameters} &  \\
			BS frequency band & $6.7$~GHz \\
			Bandwidth & $1$~MHz \\
			Number of BS sites & $3$ \\
			Total number of cells & $9$ \\
			BS transmit power & $46$~dBm \\
			BS antenna height & $25$~m \\
			Antenna type & $4\times2$ UPA \\
			Antenna directivity & Directional \\
			Electrical downtilt angle & $10^{\circ}$ \\
			\midrule
			\textbf{AN parameters} &  \\
			BS frequency band & $4.9$~GHz \\
			Bandwidth & $1$~MHz \\
			Number of BS sites & $3$ \\
			Total number of cells & $9$ \\
			BS transmit power & $46$~dBm \\
			BS antenna height & $50$~m \\
			Antenna type & $4\times2$ UPA \\
			Antenna directivity & Directional \\
			Electrical downtilt angle & $-10^{\circ}$ \\
			\midrule
			\textbf{SN parameters} &  \\
			LEO orbital altitude & $550$~km \\
			Number of LEO satellites & $2$ \\
			Satellite linear velocity & $7.5$~km/s \\
			Carrier frequency & $2.185$~GHz \\
			Bandwidth & $1$~MHz \\
			Satellite transmit power & $46$~dBm \\
			Beam pointing direction & $[0,0,-1]$ \\
			Satellite antenna configuration & $8\times8$ UPA \\
			\midrule
			\textbf{UAV parameters} &  \\
			Min flight height & $100$~m \\
			UAV speed & $5,10,15,20$~m/s \\
			UAV number & $32$ \\
			Antenna type & Omnidirectional, gain: $0$~dBi \\
			\bottomrule
		\end{tabular}
	\end{table}

	\begin{table}[t]
		\centering
		\footnotesize
		
		\captionsetup{font=scriptsize}
		
		\caption{HDRL Parameters}
		\label{HDRL Parameters}
		\setlength{\tabcolsep}{6pt}
		\begin{tabular}{p{0.55\columnwidth} p{0.37\columnwidth}}
			\toprule
			\textbf{Parameter} & \textbf{Value} \\
			\midrule
			
			\textbf{Top-Level: DDQN Parameters} &  \\
			Number of neurons in hidden layers & $128,\,64$ \\
			Exploration rate  & $0.5,\,0.05$ \\
			Replay memory and mini-batch size & $50000,\,128$ \\
			Discounting factor  & $0.97$ \\
			Target network updating period (steps) & $200$ \\
			Training network learning rate  & $0.0005$ \\
			
			\midrule
			\textbf{Lower-Level: CSAC Parameters} &  \\
			Number of neurons in hidden layers & $128,\,64$ \\
			Replay memory and mini-batch size & $50000,\,128$ \\
			Discounting factor & $0.99$ \\
			Target critic network soft updating factor & $0.005$ \\
			Actor and critic network learning rate & $0.0003$ \\
			
			\bottomrule
		\end{tabular}
		\vspace{-2mm}
	\end{table}

	\vspace{-5pt}
	
	\vspace{-5pt}
	
	\section{Numerical Results}\label{Simulation Results}


	We consider an urban area of size $2$~km $\times$ $2$~km with high-rise buildings, which represents a particularly challenging environment for coverage-aware UAV navigation.
This is because Line-of-Sight (LoS) and Non-Line-of-Sight (NLoS) links, as well as the received signal strength, may vary frequently as the UAV moves through the area~\cite{zeng2021simultaneous}.
To accurately simulate the channel between the UAV and the BS in this environment, we first generate the positions and heights of buildings based on the statistical building model recommended by the International Telecommunication Union (ITU)~\cite{ITU_R_P1410_5_2012}.
This statistical building model has been widely adopted to evaluate the LoS probability of wireless links.
Unless otherwise specified in Table~\ref{tab:main_params}, the remaining parameter settings are as follows:

	\subsubsection{GN Parameters}\label{Ground Network Topology and Links}
	
We consider a geographical area covered by three GN BS sites, each of which is sectorized into three sectors using the standard sectorization technique~\cite{3gpp36873}, resulting in a total of $9$ GN cells.
The three sectors are oriented toward azimuth angles of $0^{\circ}$, $120^{\circ}$, and $240^{\circ}$, respectively, yielding a directional radiation pattern for GN coverage~\cite{zeng2019accessing}.

	\subsubsection{AN Parameters}\label{Air Network Topology and Links}
	
The ABS is a dedicated network designed for low-altitude coverage, which is deployed at a higher altitude and employs an electrical uptilt of $10^{\circ}$~\cite{kim2022non} so that the main beam points toward the low-altitude region.
The ABS operates in frequency bands dedicated to UAV communications~\cite{kim2022non}.

	\begin{figure}[t]
		\centering		\includegraphics[width= 2.5in]{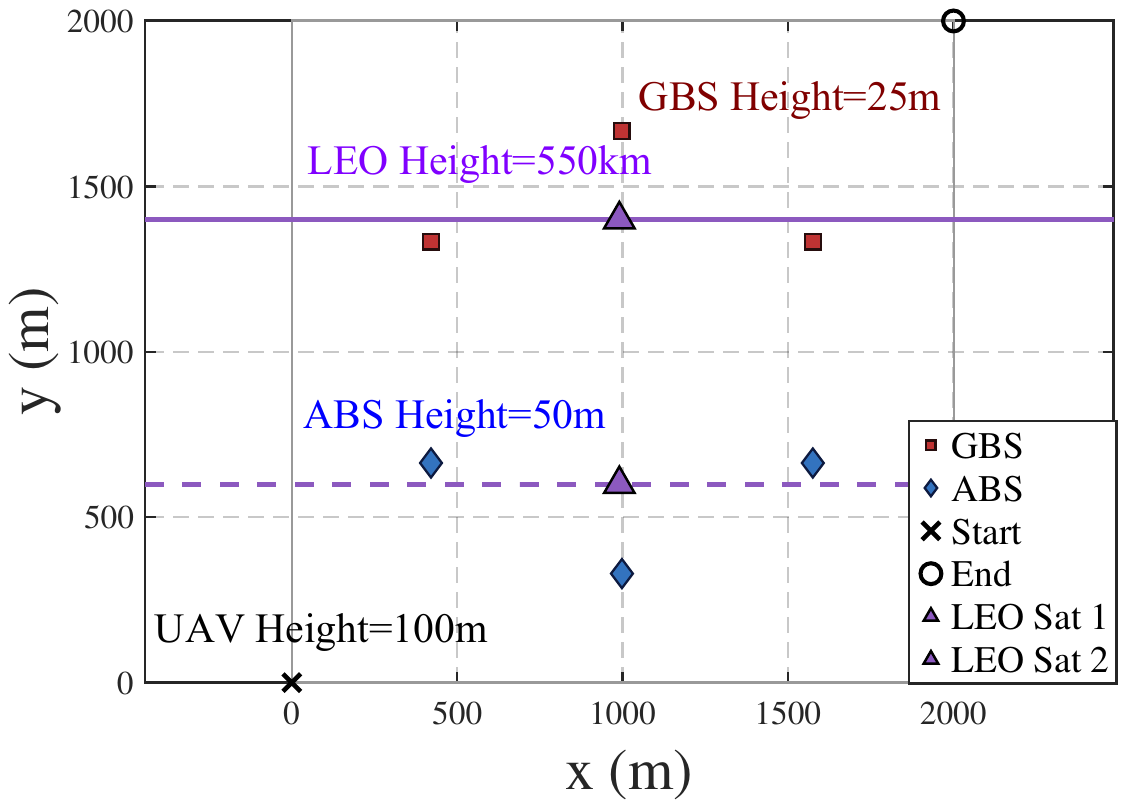}
		\caption{Illustration of the SAGIN simulation scenario.}
		\label{fig:scenario}
	\end{figure}

	\begin{figure}[t]
		\centering		\includegraphics[width= 2.5in]{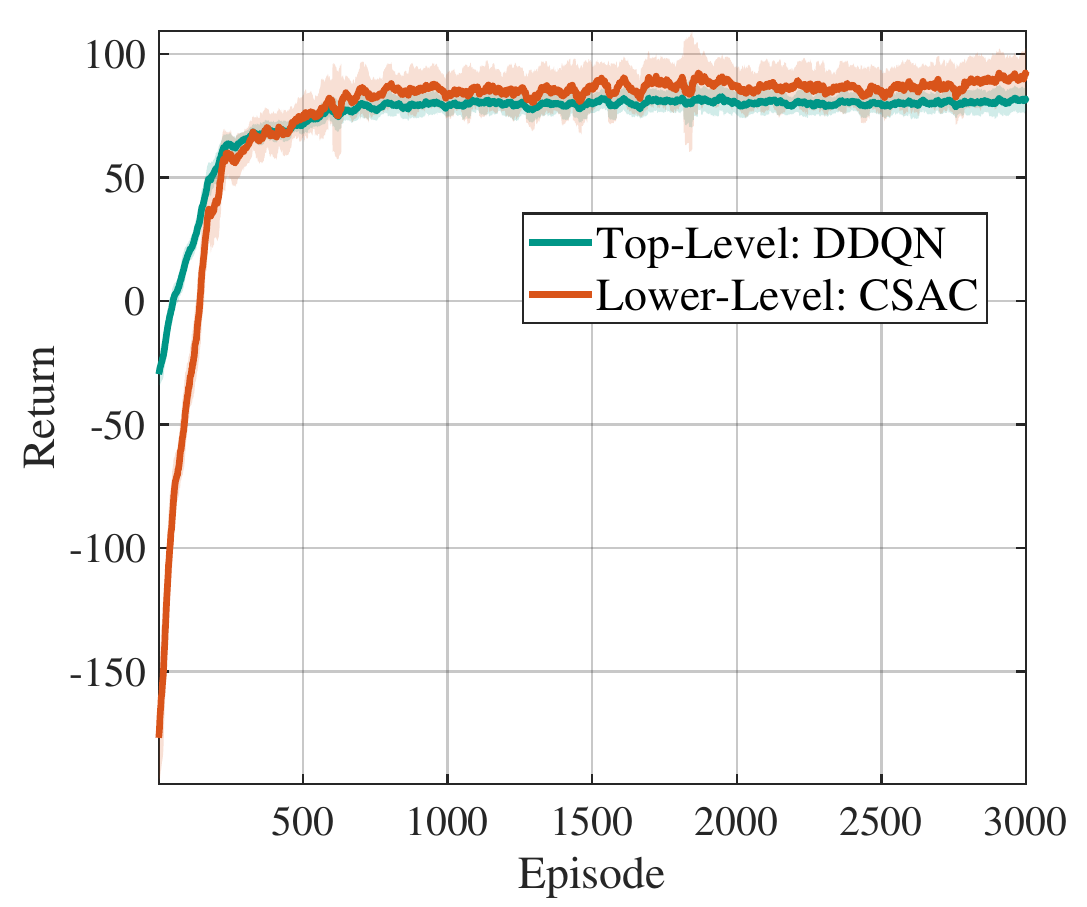}
		\caption{Illustration of the HDRL training convergence performance.}
		\label{fig:convergence}
		\vspace{-2mm}
	\end{figure}

	\subsubsection{SN Parameters}\label{Space Network Topology and Links}
    
For the SN, two LEO satellites are considered.
Both satellites move along the positive $x$-axis with a constant ground-projected velocity of 7.5~km/s~\cite{leyva2021inter}.
Each satellite employs a fixed beam pointing vertically downward and is equipped with an $8\times8$ uniform planar array (UPA)~\cite{you2024integrated}.

	\subsubsection{Links}\label{Channels}

For the AN-UAV and GN-UAV links, the building distribution generated earlier is first used to determine whether a LoS link exists between the UAV and each BS at a given UAV position.
Subsequently, the path loss is calculated according to the 3GPP urban macro (UMa) model~\cite{3gppTR36777}.
Small-scale fading is then superimposed: Rayleigh fading is adopted under NLoS conditions, while Rician fading with a Rician factor of 15~dB is applied under LoS conditions.
For the SN-UAV link, the UAV generally maintains unobstructed visibility to the LEO satellite; thus, the channel is modeled using the 3GPP NTN urban LoS scenario~\cite{3GPPTR38821}, as illustrated in Fig.~\ref{fig:scenario}.

	\subsubsection{HDRL Details}\label{HDRL Network}
	
Regarding the HDRL architecture~\cite{kulkarni2016hierarchical}, the top-level and lower-level agents are implemented as four-layer fully connected (FC) neural networks.
Activation functions are set to rectified linear units (ReLUs), and the Adam optimizer~\cite{kingma2014adam} is adopted for gradient-based optimization in the networks.
The maximum number of steps per episode is set to $300$, and the HDRL training phase consists of $3000$ episodes.
After training, the simulation enters the testing stage, during which the HDRL networks are fixed and the agent executes optimal actions based on the observed states.
Each testing phase consists of a single episode, and the detailed hyperparameter configurations are summarized in Table~\ref{HDRL Parameters}.

	\vspace{-5pt}

	
	
	\subsection{Performance Comparison}

To comprehensively evaluate the mobility management performance, we consider four metrics: the average link rate~\cite{wang2024sustainable}, the number of link selections~\cite{wang2024sustainable}, the QoS satisfaction ratio~\cite{zhang2024handover}, and the flight time~\cite{zeng2021simultaneous}.
These metrics respectively characterize the throughput performance, link switching behavior, QoS guarantee, and flight efficiency of UAVs.
Based on these metrics, we compare the proposed \textbf{DDQN+CSAC} algorithm with the following baseline methods:
\begin{itemize}
	\item \textbf{Direct RL (DRL)}~\cite{zhan2022energy}: A baseline method that applies a DDQN algorithm by discretizing the continuous action space into a finite set of actions and selects the BS with the maximum SINR for connectivity.

	\item \textbf{Straight-Line Flight (SL)}~\cite{Zhao2023LinkQualityHO}: A baseline method where the UAV flies directly from the start to the destination and connects to the BS with the highest RSRP.
	
	\item \textbf{Graph-Based}~\cite{zhang2019trajectory}: A baseline method that considers only the UAV’s connectivity with GN and AN, and uses graph theory and convex optimization to obtain feasible and near-optimal solutions.
\end{itemize}

	\begin{figure*}[t]
		\centering
		
		\begin{subfigure}[t]{0.29\textwidth}
			\centering
			\includegraphics[width=\linewidth]{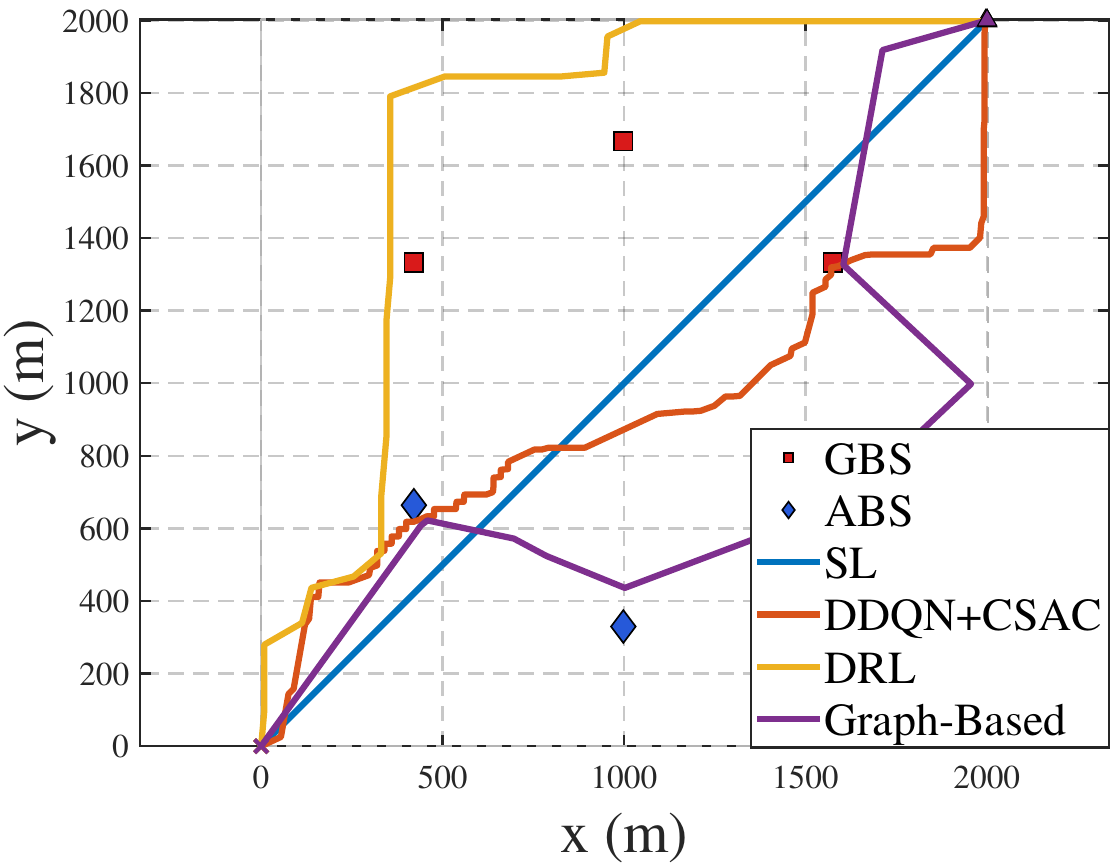}
			\caption{Trajectories.}
			\label{fig_Trajectory}
		\end{subfigure}\hfill
		\begin{subfigure}[t]{0.32\textwidth}
			\centering
			\includegraphics[width=\linewidth]{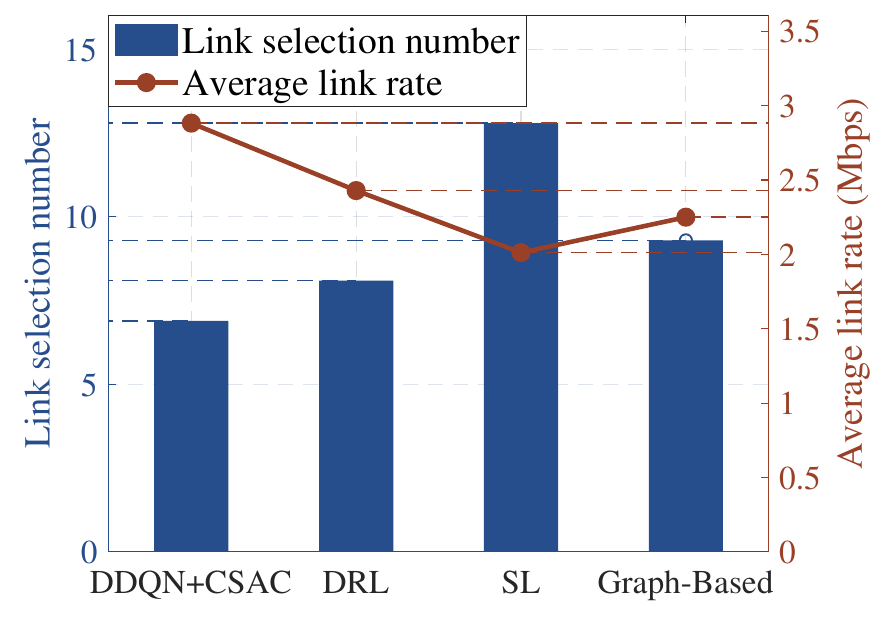}
			\caption{Link selection number and average link rate.}
			\label{fig_1UAV_RateAndHandover}
		\end{subfigure}\hfill
		\begin{subfigure}[t]{0.32\textwidth}
			\centering
			\includegraphics[width=\linewidth]{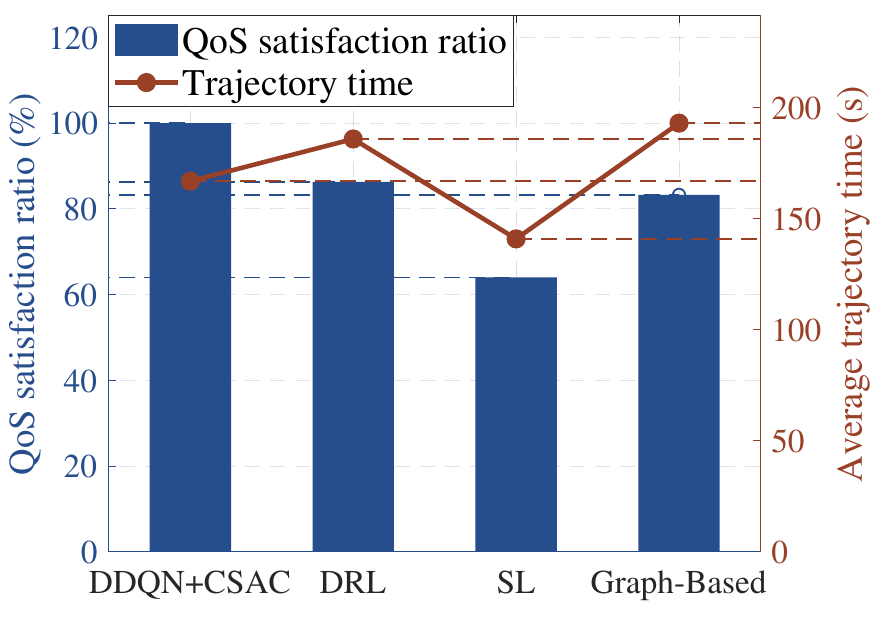}
			\caption{QoS satisfaction ratio and flight time.}
			\label{fig_1UAV_QosAndTime}
		\end{subfigure}
		
		{\captionsetup{justification=raggedright,singlelinecheck=false}
			\caption{The performance with different benchmark methods.}
			\label{fig_1UAV}}
	\end{figure*}

	\begin{figure*}[t]
		\centering
		
		\begin{subfigure}[t]{0.29\textwidth}
			\centering
			\includegraphics[width=\linewidth]{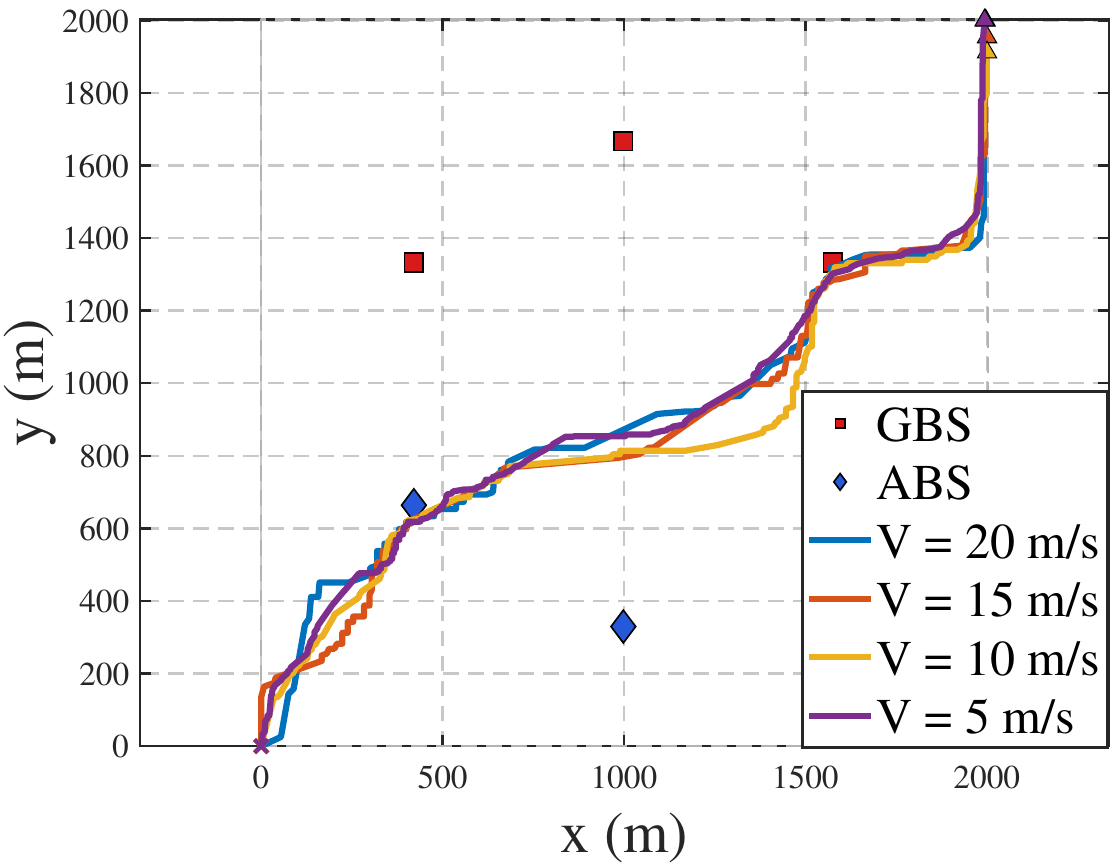}
			\caption{Trajectories.}
			\label{fig_DifferentSpeed_Trajectory}
		\end{subfigure}\hfill
		\begin{subfigure}[t]{0.32\textwidth}
			\centering
			\includegraphics[width=\linewidth]{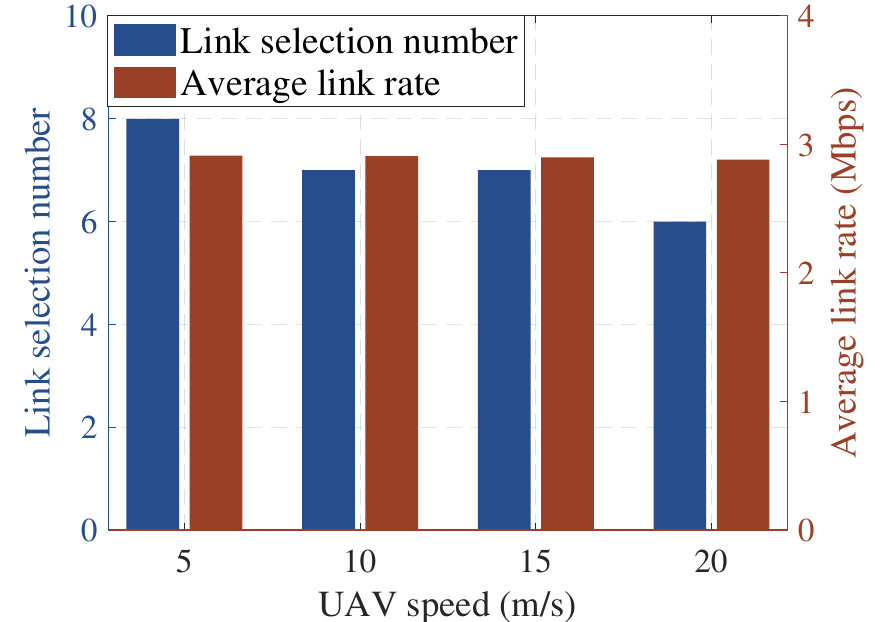}
			\caption{Link selection number and average link rate.}
			\label{fig_DifferentSpeed_RateAndHandover}
		\end{subfigure}\hfill
		\begin{subfigure}[t]{0.32\textwidth}
			\centering
			\includegraphics[width=\linewidth]{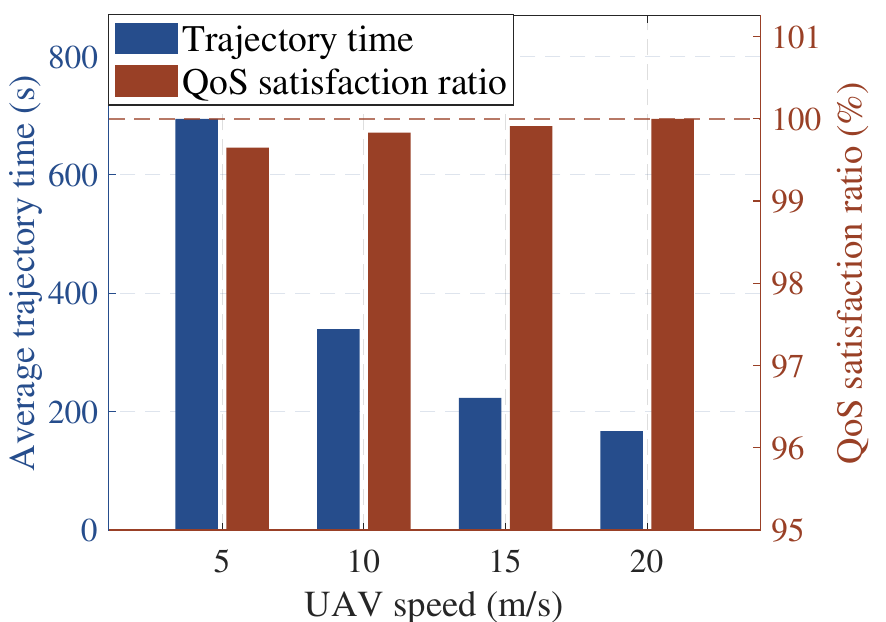}
			\caption{QoS satisfaction ratio and flight time.}
			\label{fig_DifferentSpeed_QosAndTime}
		\end{subfigure}
		
		{\captionsetup{justification=raggedright,singlelinecheck=false}
			\caption{The performance with different UAV speeds.}
			\label{fig_DifferentSpeed}}
		\vspace{-2mm}
	\end{figure*}

	Fig.~\ref{fig:convergence} illustrates the convergence behavior of the proposed \textbf{DDQN+CSAC} algorithm during training.
It can be observed that the returns of both the top-level and lower-level agents gradually converge to stable values, indicating that the proposed algorithm achieves fast convergence, stable training, and effective learning.

	Fig.~\ref{fig_Trajectory} presents the UAV flight trajectories under different algorithms, while the performance of the four key metrics is illustrated in Fig.~\ref{fig_1UAV_RateAndHandover} and Fig.~\ref{fig_1UAV_QosAndTime}.
As shown in Fig.~\ref{fig_1UAV_RateAndHandover}, the proposed \textbf{DDQN+CSAC} algorithm achieves the best performance in terms of average link rate and the number of link selections, providing gains of 18.66\% and 28.10\% compared with the \textbf{DRL} and \textbf{Graph-Based} schemes, respectively.
This performance improvement arises because the \textbf{DRL} baseline discretizes the continuous action space, leading to quantization errors, whereas the proposed \textbf{DDQN+CSAC} framework decouples discrete and continuous decision variables in Section~\ref{HDRL Framework} and employs CSAC in Section~\ref{sec:SAC} to effectively handle continuous trajectory control.
Fig.~\ref{fig_1UAV_QosAndTime} further shows that the proposed algorithm achieves the second-shortest flight time, only slightly longer than the \textbf{SL} scheme, while maintaining a 100\% QoS satisfaction ratio throughout the entire flight.
In contrast, the \textbf{DRL} and \textbf{SL} schemes achieve QoS satisfaction ratios of only 86.3\% and 64\%, respectively.
This is because conventional \textbf{DRL} methods incorporate constraints into the reward function in an indirect manner, which cannot guarantee effective constraint satisfaction during execution.
By contrast, the proposed \textbf{DDQN+CSAC} algorithm leverages CRL in Section~\ref{sec:SAC} to optimize long-term cumulative rewards while dynamically adjusting constraint weights in~\eqref{eq:constraint_rl}, thereby significantly enhancing QoS satisfaction.

	\vspace{-5pt}
	\subsection{Robustness Testing}

\begin{figure*}[!t]
		\centering
		
		\begin{subfigure}[t]{0.29\textwidth}
			\centering
			\includegraphics[width=\linewidth]{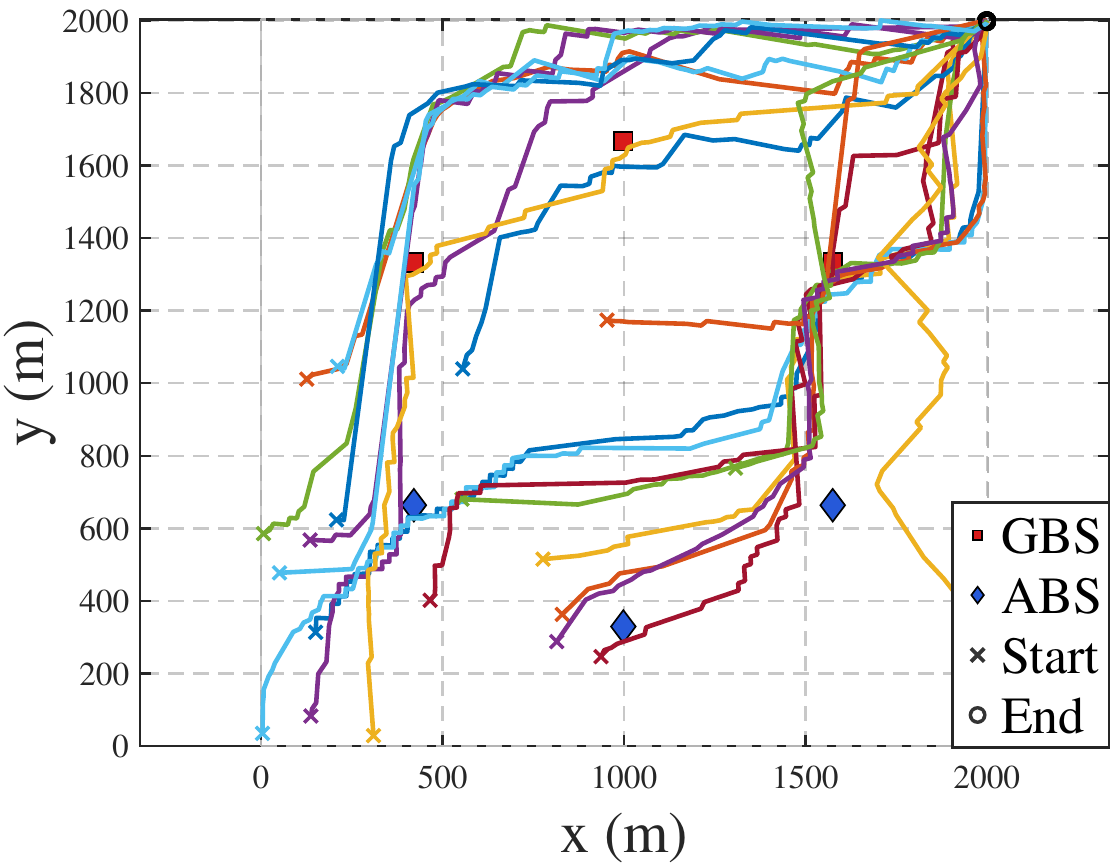}
			\caption{Trajectories.}
			\label{fig_DifferentNumber_Trajectory}
		\end{subfigure}\hfill
		\begin{subfigure}[t]{0.32\textwidth}
			\centering
			\includegraphics[width=\linewidth]{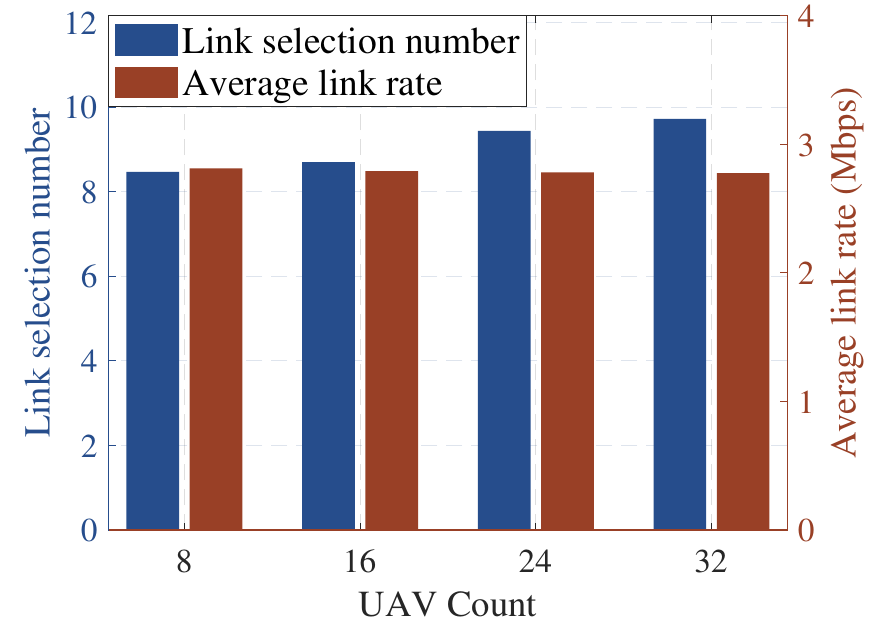}
			\caption{Link selection number and average link rate.}
			\label{fig_DifferentNumber_RateAndHandover}
		\end{subfigure}\hfill
		\begin{subfigure}[t]{0.32\textwidth}
			\centering
			\includegraphics[width=\linewidth]{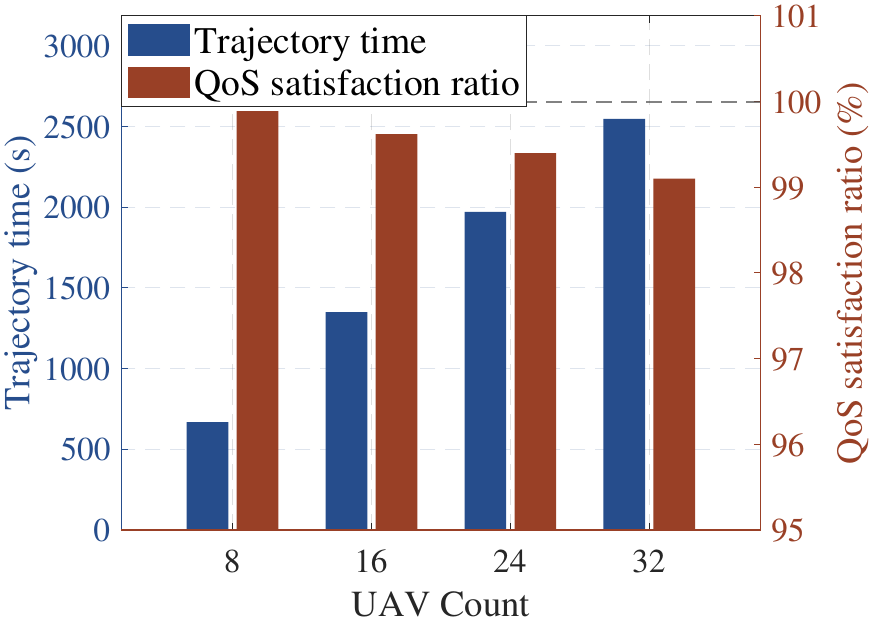}
			\caption{QoS satisfaction ratio and flight time.}
			\label{fig_DifferentNumber_QosAndTime}
		\end{subfigure}
		
		{\captionsetup{justification=raggedright,singlelinecheck=false}
			\caption{The performance with different UAV numbers.}
			\label{fig_DifferentNumber}}
		\vspace{-3mm}
	\end{figure*}

	To further evaluate the robustness of the proposed \textbf{DDQN+CSAC} framework, we vary the UAV speed and the number of UAVs to assess its performance under different  conditions.
	Fig.~\ref{fig_DifferentSpeed} illustrates the impact of different UAV velocities on the mobility management performance. 
	As illustrated in Fig.~\ref{fig_DifferentSpeed_RateAndHandover} and \ref{fig_DifferentSpeed_QosAndTime}, the total flight time decreases and the number of link selections slightly reduces as the UAV speed increases.
It can be seen that  the proposed algorithm maintains high average link rates and QoS satisfaction ratios across different UAV speeds, demonstrating strong robustness.
	In Fig.~\ref{fig_DifferentNumber},  $32$ UAVs are considered and they share the same destination location but start from different initial positions, and their corresponding flight trajectories are depicted in Fig.~\ref{fig_DifferentNumber_Trajectory}.
	In Fig.~\ref{fig_DifferentNumber_RateAndHandover} and Fig.~\ref{fig_DifferentNumber_QosAndTime}, it can be observed that as the number of UAVs increases, the average number of link selections consistently remains within $8$-$10$, the average link rate remains consistently high and stable, and the QoS satisfaction ratio remains above 99\% across all cases.
	These results demonstrate that, empowered by the CTDE mechanism, the proposed algorithm exhibits strong adaptability and stability even in multi-UAV scenarios, thereby validating its reliability and robustness.

	\begin{figure*}[!t]
		\centering
		
		\begin{subfigure}[t]{0.29\textwidth}
			\centering
			\includegraphics[width=\linewidth]{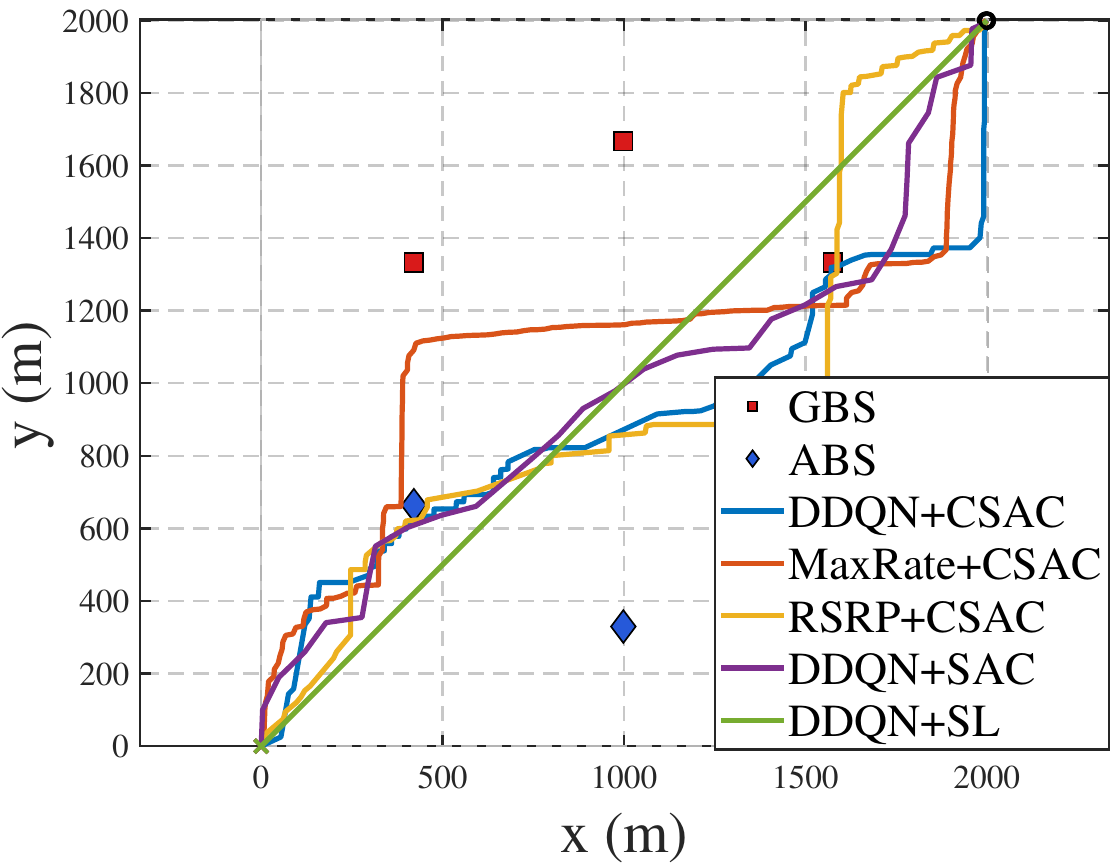}
			\caption{Trajectories.}
			\label{fig_Experiment1_Trajectory}
		\end{subfigure}\hfill
		\begin{subfigure}[t]{0.32\textwidth}
			\centering
			\includegraphics[width=\linewidth]{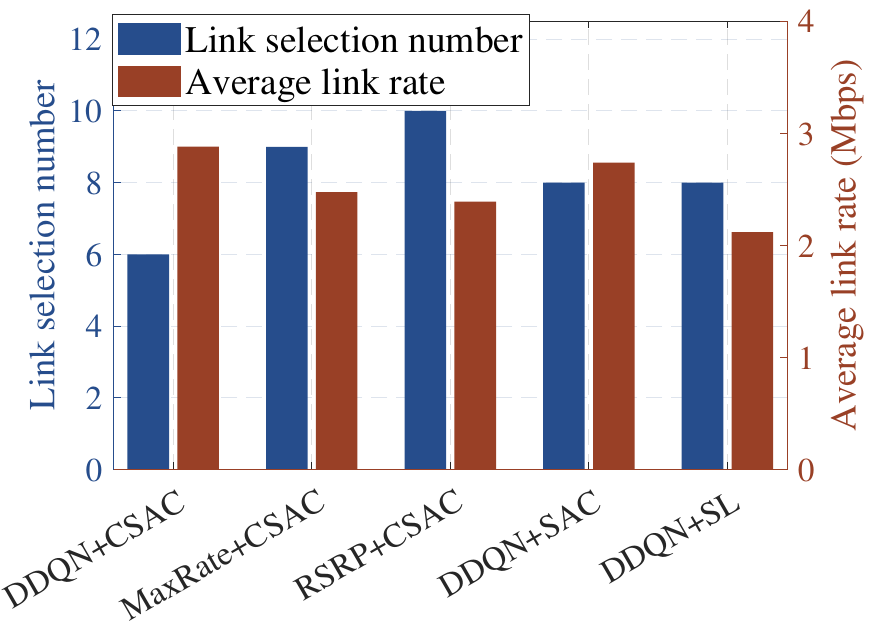}
			\caption{Link selection number and average link rate.}
			\label{fig_Experiment1_RateAndHandover}
		\end{subfigure}\hfill
		\begin{subfigure}[t]{0.32\textwidth}
			\centering
			\includegraphics[width=\linewidth]{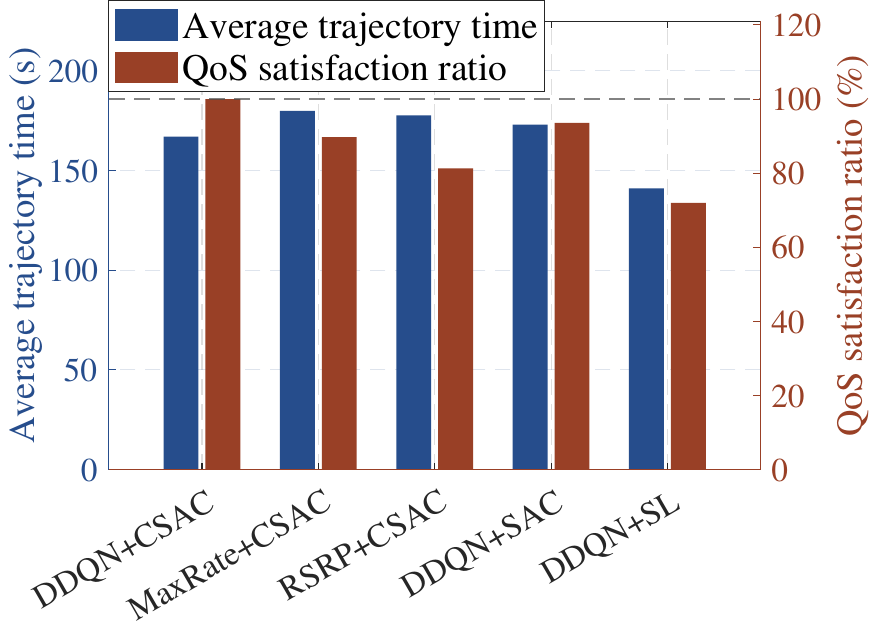}
			\caption{QoS satisfaction ratio and flight time.}
			\label{fig_Experiment1_QosAndTime}
		\end{subfigure}
		
		{\captionsetup{justification=raggedright,singlelinecheck=false}
			\caption{The performance with different ablation experiments.}
			\label{fig:number}}
		\vspace{-1mm}
	\end{figure*}

	\vspace{-1mm}
	\subsection{Ablation Experiment}
	\vspace{-1mm}
	
	To investigate how each level of HDRL
	respectively contributes to the performance gain, we perform
	an ablation experiment, where the  four
	comparing algorithms are introduced.
	{RSRP+CSAC} and {MaxRate+CSAC} adopt greedy 
	UAV-BS association strategies based on the maximum RSRP and maximum  rate in the top-level, respectively, while {DDQN+SAC} and {DDQN+SL} employ a standard SAC algorithm and a straight-line flight strategy  in the lower-level, with all remaining components identical to those of the proposed \textbf{DDQN+CSAC}.
	As shown in Fig.~\ref{fig_Experiment1_Trajectory}, \ref{fig_Experiment1_RateAndHandover}, and \ref{fig_Experiment1_QosAndTime}, both {RSRP+CSAC} and {MaxRate+CSAC} exhibit more frequent link selections and longer flight time, with their average link rates reduced by 14.09\% and 17.08\%, respectively, compared with the proposed \textbf{DDQN+CSAC}.
	This performance gap mainly arises from the DDQN long-term return learning mechanism in  Section~\ref{sec:ddqn}, which effectively balances instantaneous rate rewards and future cumulative gains, thereby reducing association-switching overhead and improving overall performance.
	In addition, although both  consider QoS constraints in the lower-level, the coupling between their top- and lower-level policies hinders unified optimization, leading to inferior results, which further demonstrates the effectiveness of the proposed DDQN in the top-level in Section~\ref{sec:ddqn}.

	Besides, the results in Fig.~\ref{fig_Experiment1_RateAndHandover} and \ref{fig_Experiment1_QosAndTime}  show that the proposed \textbf{DDQN+CSAC} achieves the fewest link selections while improving the average link rate by 5.24\% and 36.04\% compared with {DDQN+SAC} and {DDQN+SL}, respectively.
	These findings highlight the crucial role of trajectory optimization in enhancing communication performance, as jointly optimizing UAV trajectories enables better adaptation to link quality variations and significantly improves system throughput.
	The proposed \textbf{DDQN+CSAC} achieves the best performance in both flight time and QoS satisfaction rate. 
	In contrast, DDQN+SAC treats the constraints merely as penalty terms in the reward function, forming a soft-constraint mechanism that cannot guarantee strict constraint satisfaction at every time step. 
	By comparison, the proposed CRL framework in Section~\ref{sec:SAC} explicitly incorporates constraints into the optimization objective, enabling the agent to dynamically balance performance and constraint trade-offs during training, thereby achieving optimal performance while strictly satisfying all constraints, which further demonstrates the effectiveness of the proposed CSAC in the lower-level in Section~\ref{sec:SAC}.

	\vspace{-2mm}
	\section{Conclusion}\label{Conclusion}
	\vspace{-1mm}
	This paper investigated the joint link selection and trajectory optimization problem for high-mobility UAVs in SAGIN under multiple constraints. To reduce the computational complexity of traditional methods in highly dynamic environments, an HDRL framework was proposed, which decomposed the hybrid decision process into two levels: discrete UAV-BS association optimization and continuous trajectory control. The top-level employed a DDQN algorithm, while the lower-level adopted a Lagrangian-based CSAC algorithm, where both algorithms required only UAV signal measurements as input. Numerical results verified the effectiveness of the proposed HDRL framework, showing superior performance in system throughput, link selection frequency, and convergence stability compared with benchmark schemes.

	\vspace{-3mm}
	\bibliography{ref}
	\bibliographystyle{IEEEtran}

\end{document}